\documentclass[a4paper,12pt]{article}
\usepackage[utf8]{inputenc}
\usepackage{amsmath,amsthm,amsfonts,amssymb,bm}
\usepackage[right=2.5cm,left=2.5cm,top=2cm,bottom=3cm]{geometry}
\usepackage{dsfont}
\usepackage{enumitem}
\usepackage{graphicx}
\usepackage{booktabs}
\usepackage{multirow}
\usepackage{bigints}
\usepackage[authoryear]{natbib}
\usepackage[colorlinks,linkcolor=blue,citecolor=blue,urlcolor=blue]{hyperref}
\usepackage{array}
\newcolumntype{L}[1]{>{\raggedright\let\newline\\\arraybackslash\hspace{0pt}}m{#1}}
\newcolumntype{C}[1]{>{\centering\let\newline\\\arraybackslash\hspace{0pt}}m{#1}}
\newcolumntype{R}[1]{>{\raggedleft\let\newline\\\arraybackslash\hspace{0pt}}m{#1}}
\newcolumntype{P}[1]{>{\centering\arraybackslash}p{#1}}

\title{ \textsc{Variable Selection for Latent Class Analysis with Application to Low Back Pain Diagnosis}%
\thanks{This is an electronic reprint of the original article published by the The Annals of Applied Statistics, 2017, Vol. 11, No. 4, 2085--2115, DOI:10.1214/17-AOAS1061, \copyright\, Institute of Mathematical Statistics. This reprint differs from the original in pagination, typographic and other details.}}
\author{Michael Fop\thanks{School of Mathematics \& Statistics and Insight Research Centre, University College Dublin, Belfield, Dublin 4, Ireland. This work was supported by the Science Foundation Ireland funded Insight Research Centre (SFI/12/RC/2289)} \and Keith M. Smart \thanks{St. Vincent's University Hospital} \and Thomas Brendan Murphy\footnotemark[2	]}
\date{}

\begin{document}
%

\maketitle
%
%
%
%
%
%
%
%

\begin{abstract}
The identification of most relevant clinical criteria related to low back pain disorders may aid the evaluation of the nature of pain suffered in a way that usefully informs patient assessment and treatment. Data concerning low back pain can be of categorical nature, in form of check-list in which each item denotes presence or absence of a clinical condition. Latent class analysis is a model-based clustering method for multivariate categorical responses which can be applied to such data for a preliminary diagnosis of the type of pain. In this work we propose a variable selection method for latent class analysis applied to the selection of the most useful variables in detecting the group structure in the data. The method is based on the comparison of two different models and allows the discarding of those variables with no group information and those variables carrying the same information as the already selected ones. We consider a swap-stepwise algorithm where at each step the models are compared through an approximation to their Bayes factor. The method is applied to the selection of the clinical criteria most useful for the clustering of patients in different classes. 
It is shown to perform a parsimonious variable selection and to give a clustering performance comparable to the expert-based classification of patients into three classes of pain.
\end{abstract}

%

\section{Introduction}
Musculoskeletal pain is the pain concerning muscles, bones and joints, that arises in different conditions. Low back pain (LBP) is the muscoloskeletal pain related to disorders in the lumbar spine, low back muscles and nerves and it may radiate to the legs. Although there is a lack of homogeneity in the studies, a considerable proportion of the population experiences LBP during their lifetime [\cite{hoy:bain:2012,walker:2000}], with effects on social and psychic traits and working behaviour [\cite{froud:patterson:2014}]. 

Several LBP classification systems have been developed in order to group patiens into classes with similar characteristics, with the purpose of effective pain management; see \cite{stynes:2016} and references therein. Among the different systems, mechanism-based classification of pain is based on the underlying neurophysiological mechanisms responsible for its generation and maintenance. The system has been advocated in clinical practice on the ground of better pain treatment and improved patient outcomes [\cite{smart:2008,woolf:1998}]. In the absence of any diagnostic gold standards for mechanisms-based pain diagnoses, such a categorization may be identifiable on the basis of sets of symptoms and signs characteristic to each category by means of a standard clinical examination process and experienced clinical judgement [\cite{katz:2000,smart:2010,graven:2010,nijs:2015}].
Furthermore, with the aim of a diagnosis of the nature of the LBP suffered by a subject, identifying a smaller collection of signs or symptoms which best relates the manifestation of pain to its neurophysiological mechanism is a critical task. Focusing the attention only on few manifest pain characteristics can guide a preliminary patient evaluation and can constitute a valid basis for additional investigations and immediate pain treatment.

Model-based clustering [\cite{fraley:raftery:2002,mcnicholas:2016}] is a well established framework for clustering multivariate data. In this approach, the data generating process is modelled through a finite mixture of probability distributions, where each component distribution corresponds to a group. When the observations are measured on categorical variables (such as data arising from questionnaires), the most common model-based clustering method is the latent class analysis model (LCA) [\cite{lazarsfeld:henry:1968}]. Typically all the variables are considered in fitting the model, but often only a subset of the variables at hand contains the useful information about the group structure of the data. When performing variable selection for clustering the goal is to remove \emph{irrelevant} variables, which do not carry group information, and \emph{redundant} variables, which convey similar group information, retaining only with the set of \emph{relevant} variables, which contains the useful information [\cite{dy:brodley:2004}]. Therefore considering all the variables unnecessarily increases the model complexity and can produce model identifiability problems. Moreover, using variables that do not contain group information or that contain unneeded information frequently leads to a poor classification performance. 

In recent years, wide attention has been given to the problem of variable selection in clustering multivariate data. The problem has been generally tackled through two approaches: the \emph{wrapper} approach, which combines clustering and variable selection at the same time, and the \emph{filter} approach, where the variables are selected after or before the clustering is performed [\cite{dy:brodley:2004}]. Model-based clustering for continuous data has seen the prevalence of the wrapper approach; we cite the works from \citet{law:figueiredo:2004}, \citet{tadesse:sha:2005}, \citet{kim:tadesse:2006}, \citet{raftery:dean:2006}, \citet{maugis:celeux:2009:a}, \citet{maugis:celeux:2009:b}, \citet{murphy:dean:2010}, \citet{scrucca:raftery:2014}, \citet{malsiner:etal:2016}, \citet{marbac:sedki:2016}. Moreover, further works in the wrapper approach considered the introduction of a penalty term in the log-likelihood in order to induce sparsity in the features, for example \citet{pan:shen:2007}, \citet{wang:zhu:2008}, \citet{xie:pan:2008}, \citet{meynet:maugis:2012}

In LCA, the variable selection problem has been assessed only recently. Under a filter approach, \citet{zhang:ip:2014} propose two measures for quantifying the discriminative power of a variable for mixed mode data, but the method is limited only to binary variables. Under the wrapper approach,  \citet{dean:raftery:2010} recast the variable selection problem as a model selection problem, \citet{bontemps:toussile:2013} suggest an approach designed on the minimization of a risk function,  \citet{silvestre:cardoso:2015} propose a method adapted from \citet{law:figueiredo:2004} and based on feature saliency, \citet{white:wyse:2014} present a full Bayesian framework with a collapsed Gibbs sampler and \citet{bartolucci:etal:2016} present a method based on the work of \citet{dean:raftery:2010} for item selection in questionnaires. 

All of the above mentioned wrapper methods for LCA have a drawback: they consider a variable to be added or removed to the already selected set of clustering ones assuming that the former is independent of the latter. By this assumption, two (or more) informative correlated variables are selected, even if they contain similar group information. However, retaining only one (or a subset) of them can lead to a clustering of comparable quality with a more parsimonious variable selection. Thus the result is the methods are capable of discarding non informative variables, but not the redundant variables. 

In this work we develop a variable selection method for LCA based on the model selection framework of \citet{dean:raftery:2010} which overcomes the limitation of the above independence assumption. By adapting the variable role modeling of \citet{maugis:celeux:2009:b} in the variable selection procedure, we propose a method capable of discarding variables that do not contain group information and variables that are redundant. This variable selection method assesses a variable usefulness for clustering by comparing models via an approximation to their Bayes factor. 

We apply the proposed method to cluster a set of patients suffering of low back pain. Each patient were diagnosed as having a different type of pain by a group of experienced physiotherapists using a list of several clinical indicators. The aim is to recover in an unsupervised setting a classification of the patients comparable to the expert-based one and at the same time selecting a reduced collection of clinical indicators that can be used for a preliminary assessment of the characteristics of pain. 

Section~\ref{data} presents the low back pain data which gave the motivation for the improvement in the variable selection approach for LCA. In Section~\ref{LCA} we give a brief description of model-based clustering and latent class analysis. The general variable selection methodology for LCA is presented in Section~\ref{varsel}. First, we review the \citet{dean:raftery:2010} procedure and subsequently we present our proposed variable selection method characterized by the relaxation of the independence assumption between the clustering variables and the proposed one. Section~\ref{res} is dedicated to the results of the variable selection method applied to the LBP data. Section~\ref{sim} presents a simulation study on two different scenarios. The paper ends with a brief discussion in Section~\ref{disc}. 

\section{Low back pain data}
\label{data}
A mechanisms-based classification of pain relates the generation and maintenance of pain to its underlying neurophysiological mechanisms. To this purpose, the following categories have been suggested for a clinically meaningful classification of pain [\cite{merskey:bogduk:2002,woolf:2004}]:
\begin{itemize}
\item \emph{Nociceptive}: Pain that arises from actual or threatened damage to non-neural tissue, occurring with a normally functioning somatosensory nervous system;
\item \emph{Peripheral Neuropathic}: Pain initiated or caused by a primary lesion or dysfunction in the peripheral nervous system;
\item \emph{Central Sensitization}: Pain initiated or caused by a primary lesion or dysfunction in the central nervous system.
\end{itemize}

It is thought that classifying patients low back pain based on a clinical judgement regarding the likely dominant category of neurophysiological mechanisms responsible for its generation and/or persistence may usefully inform treatment by inviting clinicians to select treatments either known or hypothesized to target those mechanisms in an attempt to optimize clinical outcomes [\cite{smart:2008}]. In this regard, a list of $38$ clinical criteria (signs and symptoms) whose presence or absence can best discriminate the three types of pain has been generated on an expert-consensus basis. See \cite{smart:2010} and Supplementary Material, Section 4 [\cite{supplement}] for the complete clinical criteria checklist. 

\citet{smart:blake:2011} conducted a preliminary discriminative validity study of such mechanisms-based classification of musculoskeletal pain in clinical practice. The aim of the study was to assess the discriminative validity of the above classification system for low back disorders. The data are a sample of $464$ patients, each one assigned to one of the three categories of pain by a group of experienced physiotherapists. For each patient, information regarding the presence/absence of the $38$ binary clinical indicators is recorded.

In the present work, in analysing these data the aim is twofold:
\begin{enumerate}[noitemsep]
 \item Implement an unsupervised partition of the patients to form groups of patients with similar characteristics. Thus, we can establish if the clusters found using the unsupervised method agree with the expert-based classification or not. This allows for the discovery of a potentially novel partition of the patients into homogeneous groups or a further validation of the expert-based classification;
 \item Select a subset of most relevant clinical criteria for partitioning the patients. Most of the indicators (if not all) have good discriminative power and large part of them carry the same information about the pain categories. The interest here is to discard redundant and non-informative indicators in order to reduce the list of signs and symptoms to check for a preliminary assessment of a patient condition.
\end{enumerate}

In collecting the data, the presence/absence of some criteria was indicated as ``Don't know'' for some patients as the corresponding information was unavailable. In particular, Criteria 20 records if a subject condition was responsive or not to nonsteroidal anti-inflammatory drugs (NSAIDs) and for a set of patients it was not known if they actually took or not any NSAIDs. In \citet{smart:blake:2011} these entries were discarded. To be consistent with the authors approach and consider the same set of data we discard them as well in the following analysis; Section 1 of Supplementary Material [\cite{supplement}] contains a discussion and a brief analysis with these entries included as extra category. Furthermore, Criteria $17$ and $21$ are not available in the data and are not considered. The final data set is then composed of $425$ patients examined on $36$ binary variables.

\section{Latent class analysis}
\label{LCA}
Let $\mathbf{X}$ the $N \times M$ data matrix, where each row $\mathbf{X}_n$ is the realization of a $M$-dimensional vector of random variables $\mathbf{X}_n =  (X_{n1},\,\dots,\,X_{nm},\,\dots\,X_{nM})$. Model-based clustering assumes that each $\mathbf{X}_n$ arises from a finite mixture of $G$ probability distributions, each representing a different cluster or group. The general form of a finite mixture distribution is specified as follows:
\begin{equation}
\label{eq:1}
p(\mathbf{X}_n) = \sum_{g=1}^G \tau_g \, p(\mathbf{X}_n \lvert {\bm{\theta}}_g),
\end{equation} 
where the $\tau_g$ are the mixing probabilities and $\bm{\theta}_g$ is the parameter set corresponding to component $g$. The component densities fully characterize the group structure of the data and each observation belongs to the corresponding cluster according to a set of unobserved cluster membership indicators $\mathbf{z}_n = (z_{n1},\,z_{n2},\,\dots,z_{nG})$, such that $z_{ng} = 1$ if $\mathbf{X}_n$ arises from the $g$th subpopulation [\cite{mclachlan:peel:2000,fraley:raftery:2002}].

When clustering multivariate categorical data a common model-based approach is the latent class analysis model (LCA). In this framework, within each class each variable $X_m$ is modelled using a Multinomial distribution, therefore
\[
p(X_m \lvert \theta_g) = \prod_{c=1}^{C_m} \theta_{gmc}^{\mathds{1}\lbrace X_m = c \rbrace},
\]
where $c = 1,\,\dots,C_m$ are the possible categories values for variable $m$, $\theta_{gmc}$ is the probability of the variable taking value $c$ given class $g$, and $\mathds{1}\lbrace x_m = c \rbrace$ is the indicator function equal to 1 if the variable takes value $c$, 0 otherwise. In LCA it is assumed that the variables are statistically independent given the class value of an observation. This is a basic assumption known as \emph{local independence assumption} [\cite{clogg:1988}] and it allows the following factorization of the joint component density:
\[
p(\mathbf{X}_n \lvert \bm{\theta}_g) = \prod_{m=1}^M \prod_{c=1}^{C_m} \theta_{gmc}^{\mathds{1}\lbrace X_{nm} = c \rbrace};
\]
consequently the overall density in~\eqref{eq:1} becomes
\[
p(\mathbf{X}_n) = \sum_{g=1}^G \tau_g \prod_{m=1}^M \prod_{c=1}^{C_m} \theta_{gmc}^{\mathds{1}\lbrace X_{nm} = c \rbrace}.
\]

For a fixed value $G$ the set of parameters $\lbrace \tau_g, \theta_{\,gmc} \colon m=1,\,\dots,\,M; c=1,\,\dots,\,C_m; g=1,\,\dots,\,G \rbrace$ is usually estimated by the EM algorithm, but also a Newton-Raphson algorithm or a hybrid form of the two can be considered [\cite{mclachlan:krishnan:2008}]. In any case the algorithm is initialized through a set of randomly generated starting values and there is no guarantee of reaching the global maximum. For this reason is usually a good practice to run the procedure a number of times and select the best solution [\cite{bartholomew:knott:2011}]. 

More details about the model and the parameter estimation are provided in~\citet{lazarsfeld:henry:1968}, \citet{goodman:1974,haberman:1979}, \citet{clogg:1995}, \citet{agresti:2002} and \citet{bartholomew:knott:2011}.

Regarding parameters interpretation, in the LCA model the parameter $\theta_{gmc}$ represents the probability of occurrence of attribute $c$ for variable $X_m$ in class $g$. Thus for the binary variables of the LBP data, $\theta_{gmc}$ will represent the probability of having a certain symptom or clinical criteria for each patient belonging to class $g$.

\subsection*{Model selection}
\label{BIC}
Different LCA models are specified by assigning different values to $G$. Here the selection of the best model and of the related number of latent classes is carried out using an approximation to their Bayes factor. When comparing two competing models specified to describe the data $\mathbf{X}$, say $\mathcal{M}_A$ against $\mathcal{M}_B$, the extent to which the data support model $\mathcal{M}_A$ over $\mathcal{M}_B$ is measured by their posterior odds. In absence of prior preference for one of the two models, this quantity is given by
\[
\dfrac{p(\mathcal{M}_A \lvert \mathbf{X})}{p(\mathcal{M}_B \lvert \mathbf{X} )} = \dfrac{p(\mathbf{X} \lvert \mathcal{M}_A)}{p(\mathbf{X} \lvert \mathcal{M}_B)},
\]
where $p(\mathbf{X} \lvert \mathcal{M}_A)={\bigintss} p(\mathbf{X}\lvert\bm{\theta},\mathcal{M}_A)\,p(\bm{\theta}\lvert\mathcal{M}_A)\,d\bm{\theta}$ is the integrated likelihood. The ratio of the integrated likelihoods of the two models is the Bayes factor, $\mathcal{B}_{A,B}$. The quantity $p(\mathbf{X} \lvert \mathcal{M}_A)$ is conveniently approximated using the Bayesian Information Criterion (BIC), defined by
\[
\text{BIC}\bigl(\mathbf{X}\lvert\mathcal{M}_A\bigr) = 2\,\log(L_A^*) - \nu_A\,\log(N),
\]
where $L_A^*$ is the maximized likelihood and $\nu_A$ is the number of model parameters [\cite{schwarz:1978}]. Then the following approximation to twice the logarithm of the Bayes factor holds [\cite{kass:raftery:1995}]:
\begin{equation}
\label{eq:2}
2\,\log(\mathcal{B}_{A,B}) \approx \text{BIC}\bigl(\mathbf{X}\lvert\mathcal{M}_A\bigr) - \text{BIC}\bigl(\mathbf{X}\lvert\mathcal{M}_B\bigr),
\end{equation} 
and if this difference is greater than zero the evidence is in favour of model $\mathcal{M}_A$, otherwise in favour of $\mathcal{M}_B$. Several arguments in favor of BIC for model selection in mixture models have been given in the literature; see \citet{mclachlan:rathnayake:2014} for a recent review.

For a given number of variables, not all the models specified by assigning different values to $G$ are identifiable. In fact a necessary (though not sufficient) condition to the identifiability of a model with $G$ latent classes is
\begin{equation}
\label{eq:3}
\prod_{m=1}^M C_m > \Biggl(\, \sum_{m=1}^M C_m - M + 1\Biggr)G,
\end{equation}
with $C_m$ the number of categories taken by variable $X_m$ [\cite{goodman:1974}]. Thus when selecting the number of classes, hereafter we will consider values of $G$ for which this identifiability condition holds.


\section{Variable selection for latent class analysis} \label{varsel}
\label{deanraftery}
\begin{figure}
\centering
\makebox{\includegraphics{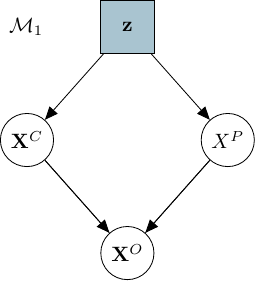}\qquad \includegraphics{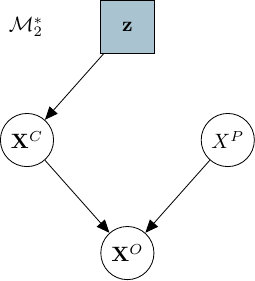}} 
\caption{\label{fig:1}The two competing models in~\citet{dean:raftery:2010}}
\end{figure}

To select the variables relevant for clustering in LCA,~\citet{dean:raftery:2010} suggested a stepwise model comparison approach. At each step of their method the authors specify a partition of the variables into
\begin{itemize}[noitemsep]
\item $\mathbf{X}^C$, the current set of relevant clustering variables, dependent on the cluster membership variable $\mathbf{z}$,
\item $X^P$, the variable proposed to be added or removed from the clustering variables,
\item $\mathbf{X}^O$, the set of the other variables which are not relevant for clustering.
\end{itemize}
Then the decision of adding or removing the considered variable is made by comparing two models: model $\mathcal{M}_1$, in which the variable is useful for clustering, and model $\mathcal{M}^*_2$ in which it does not. Figure~\ref{fig:1} gives a graphical sketch of the two competing models.

Both models make the realistic assumption that the relevant variables are not independent from the irrelevant ones (the edge between $\mathbf{X}^O$ and $\mathbf{X}^C$), but they differ in the specification of the relationship with $X^P$. In $\mathcal{M}_1$ there is no edge between $\mathbf{X}^C$ and $X^P$ because the model states that the proposed variable is useful for clustering and we have that the joint distribution $p(\mathbf{X}^C,X^P \lvert \mathbf{z})$ factorizes into $p(\mathbf{X}^C \lvert \mathbf{z})\,p(X^P \lvert \mathbf{z})$ by the local independence assumption of LCA. In $\mathcal{M}^*_2$ there is no edge between $\mathbf{z}$ and $X^P$ because under this model the proposed variable is not useful for clustering. Also, the edge between $\mathbf{X}^C$ and $X^P$ is missing, since \citet{dean:raftery:2010} assume the \emph{independence} of the proposed variable from $\mathbf{X}^C$, even when it is not relevant for clustering. However, this assumption seems to be misleading for two reasons. On one hand because if model $\mathcal{M}_2$ holds, actually $X^P$ \emph{belongs} to $\mathbf{X}^O$, contradicting the fact that the latter is not assumed independent from $\mathbf{X}^C$. On the other hand because to this assumption the model does not take into account that the proposed variable could be redundant for clustering given the set of already selected relevant variables. In fact, as it has already pointed out in previous works [\cite{law:figueiredo:2004,raftery:dean:2006,white:wyse:2014}], assuming the independence between the proposed variable and the current set of clustering variables can wrongly lead to declare as relevant a variable that could be explained by (some or all) the variables in $\mathbf{X}^C$, even if actually it contains redundant group information that is no needed or it does not contain further information at all.

\begin{figure}[!t]
\centering
\makebox{\includegraphics{fig_1a.pdf} \qquad \includegraphics{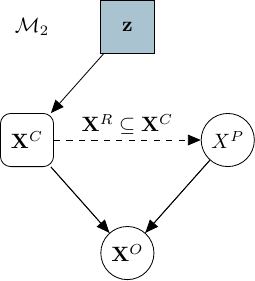}}
\caption{\label{fig:2} The two competing models specified relaxing the independence assumption between $X^P$ and $\mathbf{X}^C$.}
\end{figure}

\subsection{Relaxing the independence assumption}
\label{relaxing}
Now let us consider the models depicted in Figure~\ref{fig:2}. Model $\mathcal{M}_1$ is exactly the same model as before, where the proposed variable is useful for clustering. On the other hand, $\mathcal{M}_2$ is the model in which the proposed variable is not relevant for clustering, but there is an edge between $X^P$ and $\mathbf{X}^C$ which defines the conditional distribution $p(X^P \lvert \mathbf{X}^C)$. Therefore, $\mathcal{M}_2$ is specified by relaxing the independence assumption between the proposed variable and the set $\mathbf{X}^C$ and taking into account the potential redundancy of $X^P$. Hence, if the evidence is in favor of model $\mathcal{M}_2$, the proposed variable is discarded from $\mathbf{X}^C$ for two reasons: because it does not contain information about the latent classes at all, or because it does not add further useful information about the groups given the information already contained in the current clustering variables.

Moreover, another assumption is considered in model $\mathcal{M}_2$: we let the proposed variable to be related only to a subset $\mathbf{X}^R$ contained in the current set of clustering variables, since could be the case that not all the variables in $\mathbf{X}^C$ are associated to $X^P$ [\cite{maugis:celeux:2009:a, maugis:celeux:2009:b}].  In this way we do not induce spurious dependencies, avoiding the inclusion in the model of additional parameters without effectively increasing its likelihood. In addition a more realistic modeling framework for the relationship between $X^P$ and $\mathbf{X}^C$ is outlined, letting it to be as much flexible as possible. Clearly it ranges between two extrema: if $\mathbf{X}^R=\mathbf{X}^C$, all the current clustering variables explain $X^P$, which could likely be redundant for clustering; if $\mathbf{X}^R=\varnothing$, the proposed variable is not related to the current clustering set, recalling the assumption of \citet{dean:raftery:2010}.

Hence the two models are specified as follows:
\begin{eqnarray}
\mathcal{M}_1 \,\colon\,\, p(\mathbf{X} \lvert \mathbf{z}) &=& p(\mathbf{X}^C, X^P, \mathbf{X}^O \lvert \mathbf{z}) \nonumber\\
&=& p(\mathbf{X}^O \lvert \mathbf{X}^C, X^P)\,p(\mathbf{X}^C, X^P \lvert \mathbf{z}); \nonumber\\
\mathcal{M}_2 \,\colon\,\, p(\mathbf{X} \lvert \mathbf{z}) &=& p(\mathbf{X}^C, X^P, \mathbf{X}^O \lvert \mathbf{z})\nonumber\\
&=& p(\mathbf{X}^O \lvert \mathbf{X}^C, X^P)\,p(\mathbf{X}^C \lvert \mathbf{z})\,p(X^P \lvert \mathbf{X}^R \subseteq \mathbf{X}^C).\nonumber
\end{eqnarray}

Following \citet{dean:raftery:2010}, models $\mathcal{M}_1$ and $\mathcal{M}_2$ are then compared via the Bayes factor
\[
\mathcal{B}_{1,2}=\dfrac{p(\mathbf{X} \lvert \mathcal{M}_1)}{p(\mathbf{X} \lvert \mathcal{M}_2)}.
\]

Model $\mathcal{M}_1$ is specified by the probability distribution of the latent class model $p(\mathbf{X}^C, X^P \lvert$ $\bm{\theta}_1^C, \bm{\theta}_1^P, \mathcal{M}_1)$ and the distribution $p(\mathbf{X}^O \lvert \mathbf{X}^C, X^P, \bm{\theta}_1^{O}, \mathcal{M}_1)$. We denoted $\bm{\theta}_1^C$, $\bm{\theta}_1^P$ and $\bm{\theta}_1^O$ the parameters vectors that identify these distributions and we assume that their prior probability distributions are independent. Hence the integrated likelihood factors as follows:
\[
p(\mathbf{X}\lvert\mathcal{M}_1) = p(\mathbf{X}^O \lvert \mathbf{X}^C, X^P, \mathcal{M}_1)\,p(\mathbf{X}^C, X^P \lvert \mathcal{M}_1),
\]
with
\begin{eqnarray}
p(\mathbf{X}^O \lvert \mathbf{X}^C, X^P, \mathcal{M}_1) = \int p(\mathbf{X}^O \lvert \mathbf{X}^C, X^P, \bm{\theta}_1^{O}, \mathcal{M}_1)\,p(\bm{\theta}_1^{O}\lvert\mathcal{M}_1)\,d\bm{\theta}_1^{O};\nonumber\\
p(\mathbf{X}^C, X^P \lvert \mathcal{M}_1) = \iint p(\mathbf{X}^C, X^P \lvert \bm{\theta}_1^C, \bm{\theta}_1^P, \mathcal{M}_1)\,p(\bm{\theta}_1^C,\bm{\theta}_1^P\lvert\mathcal{M}_1)\,d\bm{\theta}_1^{C}\,d\bm{\theta}_1^{P}\nonumber.
\end{eqnarray}

Similarly the integrated likelihood of model $\mathcal{M}_2$ factors in
\[
p(\mathbf{X}\lvert\mathcal{M}_2) = p(\mathbf{X}^O \lvert \mathbf{X}^C, X^P, \mathcal{M}_2)\,p(\mathbf{X}^C \lvert \mathcal{M}_2)\,p(X^P\lvert\mathbf{X}^R\subseteq\mathbf{X}^C,\mathcal{M}_2),
\]
with
\begin{eqnarray}
p(\mathbf{X}^O \lvert \mathbf{X}^C, X^P, \mathcal{M}_2) &=& \int p(\mathbf{X}^O \lvert \mathbf{X}^C, X^P, \bm{\theta}_2^{O}, \mathcal{M}_2)\,p(\bm{\theta}_2^{O}\lvert\mathcal{M}_2)\,d\bm{\theta}_2^{O};\nonumber\\
p(\mathbf{X}^C\lvert \mathcal{M}_2) &=& \int p(\mathbf{X}^C\lvert \bm{\theta}_2^C, \mathcal{M}_2)\,p(\bm{\theta}_2^C\lvert\mathcal{M}_2)\,d\bm{\theta}_2^{C};\nonumber\\
p(X^P\lvert\mathbf{X}^R\subseteq\mathbf{X}^C,\mathcal{M}_2) &=& \int p(X^P\lvert\mathbf{X}^R\subseteq\mathbf{X}^C \bm{\theta}_2^P, \mathcal{M}_2)\,p(\bm{\theta}_2^P\lvert\mathcal{M}_2)\,d\bm{\theta}_2^{P}\nonumber.
\end{eqnarray}

Assuming that the prior distributions for $\bm{\theta}_1^{O}$ and $\bm{\theta}_2^{O}$ are the same under both models, we obtain that $p(\mathbf{X}^O \lvert \mathbf{X}^C, X^P, \mathcal{M}_1) = p(\mathbf{X}^O \lvert \mathbf{X}^C, X^P, \mathcal{M}_2)$. Therefore
\[
\mathcal{B}_{1,2} = \dfrac{p(\mathbf{X}^C, X^P \lvert \mathcal{M}_1)}{p(\mathbf{X}^C \lvert \mathcal{M}_2)\,p(X^P \lvert \mathbf{X}^R \subseteq \mathbf{X}^C, \mathcal{M}_2)}.
\]
Note that in the Bayes factor the distribution of the non-clustering variables given the rest cancels out; this represents an advantage in terms of computations because there is no need to specify the joint distribution of all the non-clustering variabels, unlike in \cite{white:wyse:2014} for example. Then this Bayes factor is estimated by the BIC approximation outlined in \eqref{eq:2}, leading to the following criterion:
\begin{eqnarray}
\text{BIC}_{\text{diff}} &=& \text{BIC}(\mathbf{X}^C,X^P\lvert\mathcal{M}_1) - \text{BIC}(\mathbf{X}^C,X^P\lvert\mathcal{M}_2)\nonumber\\
&=& \text{BIC}(\mathbf{X}^C,X^P\lvert\mathbf{z},\mathcal{M}_1) \nonumber\\
&& - \Bigl[ \text{BIC}(\mathbf{X}^C \lvert \mathbf{z},\mathcal{M}_2) + \text{BIC}(X^P\lvert\mathbf{X}^R \subseteq \mathbf{X}^C, \mathcal{M}_2 ) \Bigr]\nonumber,
\end{eqnarray}
where $\text{BIC}(\mathbf{X}^C,X^P\lvert\mathbf{z}, \mathcal{M}_1)$ and $\text{BIC}(\mathbf{X}^C \lvert \mathbf{z}, \mathcal{M}_2)$ are the BIC of the LCA model on the sets $\mathbf{X}^C\cup X^P$ and $\mathbf{X}^C$ respectively, while $\text{BIC}(X^P\lvert\mathbf{X}^R \subseteq \mathbf{X}^C, \mathcal{M}_2 )$ is the BIC of the model for the conditional distribution of the proposed variable (note that we made explicit the dependence on the latent variable $\mathbf{z}$).
If this difference is greater than zero, there is evidence in favor of $X^P$ adding further information about the clusters to the information already contained in the current set $\mathbf{X}^C$. On the other hand, if the difference is less than zero there is evidence that no useful information is added by the proposed variable.

\subsection{Proposed variable conditional distribution}
The conditional distribution of the proposed variable given $\mathbf{X}^C$ is modeled by a multinomial logistic regression using the \emph{softmax} link function:
\begin{equation}
\label{eq:4}
p(X^P = c \lvert \mathbf{X}^R \subseteq \mathbf{X}^C) = \dfrac{e^{\mathbf{X}^R\boldsymbol{\beta}_c}}{\sum_{c=1}^{C_P} e^{\mathbf{X}^R\boldsymbol{\beta}_c}},
\end{equation}
where $\boldsymbol{\beta}_c$ is the vector of regression parameters for category $c$ and $c = 1,\,\dots,\,C_P$ are the categories for the proposed variable; the model reduces to a standard logistic regression with logit link if the proposed variable is binary. We refer to \citet{ripley:1996} and \citet{agresti:2002} for a detailed description of the model and its estimation.

In the regression model \eqref{eq:4} the subset $\mathbf{X}^R$ contains the relevant predictors of the proposed variable. Their selection is carried out using a standard stepwise algorithm described in the Supplementary Material, Section 2 [\cite{supplement}].  When selecting the variables that compose $\mathbf{X}^R$, we allow it to be the empty set, thus taking into account the general variable role modeling described in \citet{maugis:celeux:2009:b}.

If the proposed variable is highly correlated with the predictors, the problem of \emph{separation} may occur. Separation arises when a linear combination of the predictors perfectly or quasi-perfectly separates the classes of the response variable, leading to infinite estimates of the regression coefficients and large standard errors [\cite{albert:anderson:1984,lesaffre:albert:1989}].
Different remedies have been proposed in literature in order to perform inference on the parameters, for example \citet{heinze:schemper:2003}, \cite{zorn:2005} and \citet{gelman:etal:2008}. 
In the present framework separation does not represent a problem, as the regression coefficients are only accessory to the computation of the maximum of the log-likelihood of the logistic regression. In fact, even in case of separation the log-likelihood surface is concave, bounded above and has a finite maximum [\cite{albert:anderson:1984}]. In practice, if separation occurs the log-likelihood surface becomes flat, approaching a limiting value as some (or all) regression coefficients are going to infinity. So convergence criteria are satisfied, and the log-likelihood is numerically maximized and computation of quantities based on that maximum, such as the BIC, are still valid [\cite{agresti:2015,albert:anderson:1984}].

\subsection{Swap-stepwise selection algorithm}
\label{algorithm}
The clustering variables are selected using a stepwise algorithm which alternates between exclusion, inclusion and swapping steps. In the removal step all the variables in $\mathbf{X}^C$ are examined in turn to be removed from the set. In the inclusion step all the variables in $\mathbf{X}^O$ are examined in turn to be added to the clustering set. In the swapping step, a non-clustering variable is swapped with a clustering variable.

In the removal and inclusion step we compare model $\mathcal{M}_1$ against model $\mathcal{M}_2$. Instead, in the swapping steps we actually compare two different configurations of model $\mathcal{M}_2$ that differ in the fact that one clustering variable is replaced by one of the non-clustering variables. The rationale for the swap step lies in the assumptions of model $\mathcal{M}_2$. In model $\mathcal{M}_2$ the proposed variable is assumed independent from $\mathbf{z}$ conditionally on the set of already selected variables and not marginally (which would be a special case). Therefore $X^P$ is actually allowed to contain some information about the clusters, which in some situations may be the best information available if one of the variables of the optimal set for $\mathbf{X}^C$ has been discarded during the search. Hence the algorithm could converge to a sub-optimum. To avoid it we compare two different sets of clustering variables in the swapping step. Then if a ``true'' clustering variable has been removed during the search, when compared to a less informative one is likely to be added back to the clustering set. 

The algorithm also performs the selection of the number $G$ of latent classes, finding at each stage the optimal combination of clustering variables and number of classes. The procedure stops when no change has been made to the set $\mathbf{X}^C$ after consecutive exclusion, swapping, inclusion and swapping steps.

A detailed description of the algorithm is in Appendix \ref{app:algo}.

\subsection{Comparing selected and discarded variables}
\label{comparing}
By means of the outlined variable selection procedure we aim to remove variables that do not contain any information about the clustering \emph{and} variables that contain \emph{additional} information, which are redundant given the already selected relevant variables. Since it is likely that related variables carry similar information about the groups, it is of interest to analyze the association between each discarded variable and each selected one after the selection is performed. We accomplish this task as a result of simple considerations.

Let $X_o \in \mathbf{X}^O$ be one of the discarded variables, and $X_c \in \mathbf{X}^C$ be one of the selected ones; let also $\hat{\mathbf{z}}$ be the estimated cluster membership allocation vector. In the light of the described general modeling framework, we analyze the association between $X_c$ and $X_o$ by comparing the following two models for the joint conditional distribution $p(X_c\,,X_o\lvert\hat{\mathbf{z}})$:
\begin{eqnarray}
\mathcal{M}_\text{as} & \colon & p(X_c\,,X_o\lvert\hat{\mathbf{z}}) = p(X_c\lvert\hat{\mathbf{z}})p(X_o\lvert X_c);\nonumber\\ 
\mathcal{M}_\text{no as} & \colon & p(X_c\,,X_o\lvert\hat{\mathbf{z}}) = p(X_c\lvert\hat{\mathbf{z}})p(X_o).\nonumber
\end{eqnarray}
In a similar fashion to the models involved in the variable selection procedure, this two models are compared via the Bayes factor $\mathcal{B}_{\text{as, no as}}=p(X_c\,,X_o\lvert\hat{\mathbf{z}}\,, \mathcal{M}_\text{as}) / p(X_c\,,X_o\lvert\hat{\mathbf{z}}\,, \mathcal{M}_\text{no as})$. Applying the same arguments of Section~\ref{relaxing} and noting that $p(X_c\lvert\hat{\mathbf{z}}\,, \mathcal{M}_\text{as}) = p(X_c\lvert\hat{\mathbf{z}}\,, \mathcal{M}_\text{no as})$ we obtain that the above Bayes factor reduces to
\[
\mathcal{B}_{\text{as, no as}} = \dfrac{p(X_o\lvert X_c\,,\mathcal{M}_\text{as})}{p(X_o \lvert \mathcal{M}_\text{no as})}.
\]
Then using the BIC approximation of \eqref{eq:2} leads to
\begin{equation}
\label{eq:5}
\mathcal{B}_{\text{as, no as}} \approx \text{BIC}_{\text{diff as}} = \text{BIC}(X_o\lvert X_c, \mathcal{M}_\text{as}) - \text{BIC}(X_o \lvert \mathcal{M}_\text{no as}).
\end{equation}
The quantity $\text{BIC}_{\text{diff as}}$ corresponds to the difference between the BIC of a multinomial logistic regression where $X_o$ depends on $X_c$ and the BIC of the regression with only the constant terms. Then if this difference is greater than zero, there is evidence of the association between the considered selected variable and the discarded one.

\section{Latent class model and clinical criteria selection}
\label{res}
The proposed model is applied to the low back pain data. We measure the agreement between the model-based partition of the data and the expert-based classification using the adjusted Rand index (ARI) which is equal to 1 when two partitions are exactly the same, otherwise it is close to 0 when they do not agree [\cite{hubert:arabie:1985}]; compared to other indices, \citet{milligan:cooper:1986} recommended the ARI as the index of choice for clustering validation.   

We consider LCA models with the number of latent classes $G$ ranging from 1 to 7. The clustering results for the different models are summarized in Table~\ref{tab1}. When fitting a LCA model on all of the clinical criteria, the BIC selects a model with 5 classes, providing an ARI of 0.50. By fixing the number of classes equal to 3 in advance we obtain a model with an ARI of 0.82. By performing the variable selection with the independence assumption of \citet{dean:raftery:2010} only one variable is discarded, Criterion 36,  and the BIC selects again a model with 5 classes, identifying the same clusters of the model on all the variables. Note that also in \citet{white:wyse:2014} only one variable is discarded. Using the variable selection method proposed here with swap-stepwise search we retain 11 variables and the BIC selects a 3-class model on these. The ARI for the model on the 11 selected clinical criteria is 0.75, thus the number of variables is reduced by about two thirds, identifying a partition of the patients that agrees well with the physiotherapists' classification. For comparison we also performed the same variable selection with a standard stepwise search, selecting a model on 10 criteria, but with a smaller ARI. Therefore the use of the swap move in the search avoided selection of sub-optimal informative clustering variables.

\begin{table}
\centering
\caption{\label{tab1} Clustering summary of the LCA model for different sets of variables and different number of classes for the LBP data (note that the BIC are not comparable for differing sets of variables).}
\begin{tabular}{llcrr}
\toprule
\begin{tabular}{@{}l@{}}\textbf{Selection}\\\textbf{method}\end{tabular} & \textbf{Variables}	&\begin{tabular}{@{}l@{}}\textbf{N. latent}\\\textbf{classes}\end{tabular} & \textbf{BIC} & \textbf{ARI}\\
\midrule
--	&	All		& 5		& -12586.48	& 0.50\\
--	&	All		& $3^*$		& -12763.81	& 0.82\\
Dean and Raftery	&	35 Criteria	& 5		& -12116.32	& 0.50\\
Stepwise		&	10 Criteria	& 3		& -3462.82	& 0.66\\
Swap-stepwise		&	11 Criteria	& 3		& -3946.31	& 0.75\\
\bottomrule
\end{tabular}\\
{\footnotesize{$^*$ We fixed the number of classes to this value in advance.}}
\end{table}

\begin{table}
\caption{\label{crosstab1} Cross-tabulation between the estimated partition on the 11 clustering variables and the expert-based classification of the LBP data.}
\centering
\begin{tabular}{llccc}
\toprule
& & \multicolumn{3}{c}{\textbf{\emph{Estimated}}}\\
& & \textbf{Class 1} & \textbf{Class 2} & \textbf{Class 3} \\
\multirow{3}{4em}{\emph{\textbf{Expert-based}}} & \textbf{Nociceptive} & 210 &   21 &  4 \\ 
& \textbf{Peripheral Neuropathic} &   5 &   88 &  2 \\ 
& \textbf{Central Sensitization} &   3 &  3 &   89 \\ 
\bottomrule
\end{tabular}
\end{table}

A cross-tabulation of the estimated partition on the 11 selected variables versus the expert-based classification is reported in Table~\ref{crosstab1}. It seems reasonable to match the 3 detected classes to the Nociceptive, Peripheral Neuropathic and Central Sensitization group respectively. 

\begin{figure}[!b]
\centering
\makebox{\includegraphics[scale=0.71]{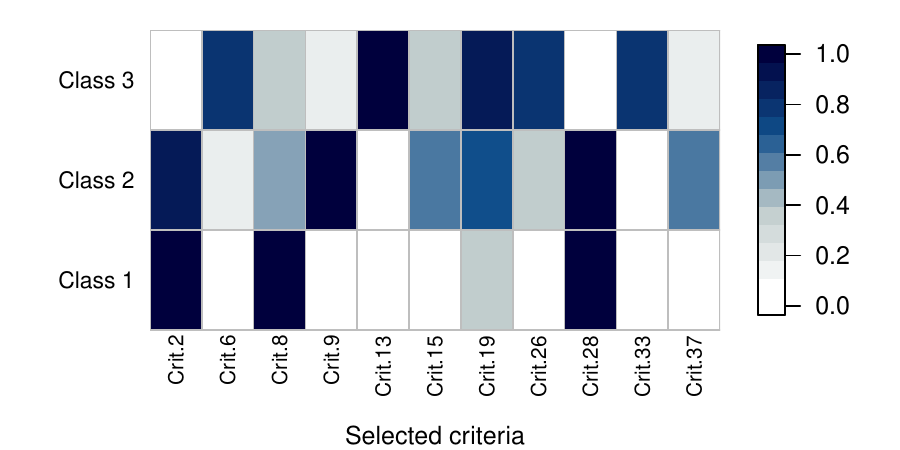}}
\caption{\label{fig:3} Heatmap plot of the estimated probability of occurrence of the 11 selected clinical criteria.}
\end{figure}

Table~\ref{tab2} lists the 11 selected clinical criteria and the estimated probability of occurrence given the class which a patient is assigned to; also the observed proportion of occurrence is reported in brackets. Figure~\ref{fig:3} is a heatmap of the estimated class conditional probabilities: the selected variables present good degree of separation between the three classes which are generally characterized by the almost full presence or almost complete absence of the selected criteria.

\begin{table}[!t]
\caption{\label{tab2} Estimated class conditional probability of occurrence and actual frequency (in brackets) for the selected clinical criteria in the low back pain data.}
\centering
\renewcommand{\arraystretch}{1.2}
\begin{tabular}{P{0.8cm}L{6cm}C{1.2cm}C{1.2cm}C{1.2cm}}
\toprule
\em \textbf{\emph{Crit.}} & \em \textbf{\emph{Description}} & \em \textbf{\emph{Class 1}} & \em \textbf{\emph{Class 2}} & \em \textbf{\emph{Class 3}} \\ 
\midrule
2  & \footnotesize{Pain associated to trauma, pathologic process or dysfunction}& %
							0.94\newline[-4pt] (0.94) & 0.90\newline[-4pt] (0.92) & 0.04\newline[-4pt] (0.04) \\
6  & \footnotesize{More constant/unremitting pain}& %
							0.05\newline[-4pt] (0.04) & 0.13\newline[-4pt] (0.17) & 0.79\newline[-4pt] (0.79) \\
8  & \footnotesize{Pain localized to the area of injury/dysfunction}& %
							0.97\newline[-4pt] (0.97) & 0.50\newline[-4pt] (0.42) & 0.31\newline[-4pt] (0.33) \\
9  & \footnotesize{Pain referred in a dermatomal or cutaneous distribution}& %
							0.06\newline[-4pt] (0.12) & 1.00\newline[-4pt] (0.97) & 0.11\newline[-4pt] (0.13) \\
13  & \footnotesize{Disproportionate, nonmechanical, unpredictable pattern of pain}& %
							0.01\newline[-4pt] (0.01) & 0.00\newline[-4pt] (0.01) & 0.91\newline[-4pt] (0.87) \\
15  & \footnotesize{Pain in association with other dysesthesias}& %
							0.03\newline[-4pt] (0.06) & 0.51\newline[-4pt] (0.51) & 0.34\newline[-4pt] (0.34) \\
19  & \footnotesize{Night pain/disturbed sleep}& %
							0.34\newline[-4pt] (0.37) & 0.70\newline[-4pt] (0.68) & 0.86\newline[-4pt] (0.85) \\
26  & \footnotesize{Pain in association with high levels of functional disability}& %
							0.07\newline[-4pt] (0.09) & 0.36\newline[-4pt] (0.36) & 0.79\newline[-4pt] (0.78) \\
28  & \footnotesize{Clear, consistent and proportionate pattern of pain}& %
							0.97\newline[-4pt] (0.95) & 0.94\newline[-4pt] (0.94) & 0.07\newline[-4pt] (0.12) \\
33  & \footnotesize{Diffuse/nonanatomic areas of pain/tenderness on palpation}& %
							0.03\newline[-4pt] (0.03) & 0.01\newline[-4pt] (0.01) & 0.73\newline[-4pt] (0.73) \\
37  & \footnotesize{Pain/symptom provocation on palpation of relevant neural tissues}& %
							0.07\newline[-4pt] (0.09) & 0.57\newline[-4pt] (0.58) & 0.19\newline[-4pt] (0.21) \\
\bottomrule
\end{tabular}
\end{table}

\citet{smart:blake:2011} fit a logistic regression of each type of pain versus the others, ending with the selection of a set of 14 features whose presence or absence best describes each class of pain. They selected Criteria 3, 4, 5, 7, 8, 9, 11, 13, 15, 19, 25, 27, 29, 33. Six out of eleven of our selected criteria match those selected  in a supervised setting. Furthermore the estimated parameters reported in Table~\ref{tab2} agree with the description of the factors related to each class of pain given by the authors: Nociceptive pain (Class 1) is well described by the presence of a pain localized to the area of injury or dysfunction, and by the absence of dysaesthesias (unpleasant sensations, e.g. crawling) and pain at night; Peripheral Neuropathic (Class 2) is characterized by the presence of a dermatomal distribution of pain and pain on palpation of nerve tissue, and Central Neuropathic (Class 3) is linked to the presence of pain that is more constant and has a disproportionate and unpredictable pattern of provocation and is associated with diffuse areas of pain on palpation as well as the absence of pain in proportion to trauma or pathology in addition to consistent and proportionate pain on clinical provocation tests. Also, fitting a LCA model on the criteria selected by \citet{smart:blake:2011}, a model with 3 latent classes is chosen, with an ARI of 0.77. By comparing the latter partition with the classification of the LCA model on the 11 criteria of Table~\ref{tab2}, an ARI of 0.79 is obtained. Thus the classification attained by the variable selection method in an unsupervised setting has a satisfying rate of agreement with the classification of patients based on the variables selected in a supervised setting, and with a smaller set of relevant clinical criteria. These findings provide some confirmatory discriminative validity evidence for a three-category mechanisms-based classification system for musculoskeletal pain.
Furthermore it is shown that the proposed method is able to reduce the number of useful clinical criteria to be checked for elaborating a preliminary assessment of the pain characteristics.

\subsection*{Discarded clinical criteria}
The clinical criteria in the data are specified in advance on a expert-consensus basis [\cite{smart:2010}]. Indeed they were chosen such that most are good in discriminating between the three types of pain. Here we want to point again the fact that the discarded criteria are removed from the set of clustering ones not only because they may not contain discriminative information about the pain classes, but also because they may carry information that is not needed, as it is already included in the set of selected ones.

We fit a LCA model on the 25 removed clinical criteria, selecting a model with 4 latent classes with a BIC of -9355.407 and an agreement to the experts' classification of 0.51. By setting in advance the number of classes equal to 3, we obtain a model with a BIC of -9470.552 and an ARI of 0.73. The cross tabulation of the fitted classification and the expert-based one for the 3-class model is presented in Table~\ref{crosstab2} 

\begin{table}
\caption{\label{crosstab2} Cross-tabulation between the estimated 3-class partition on the discarded variables and the expert-based classification of the LBP data.}
\centering
\begin{tabular}{llccc}
\toprule
& & \multicolumn{3}{c}{\textbf{\emph{Estimated}}}\\
& & \textbf{Class 1} & \textbf{Class 2} & \textbf{Class 3} \\
\multirow{3}{4em}{\emph{\textbf{Expert-based}}} & \textbf{Nociceptive} & 208 &   25 &  2 \\ 
& \textbf{Peripheral Neuropathic} &   2 &   91 &  2 \\ 
& \textbf{Central Sensitization} &   8 &  1 &   86 \\ 
\bottomrule
\end{tabular}
\end{table}

The partition thus obtained is comparable to the partition estimated on the selected clinical criteria.  Therefore by taking into consideration the discarded clinical criteria, it is still possible to get an acceptable classification of patients into clusters that sufficiently agrees with the expert-based classification. Thus it seems reasonable to consider the fact that the removed criteria are discarded mostly because they are redundant given the set of 11 selected clustering clinical criteria.

\begin{figure}[!t]
\centering
\makebox{\includegraphics[scale=0.5]{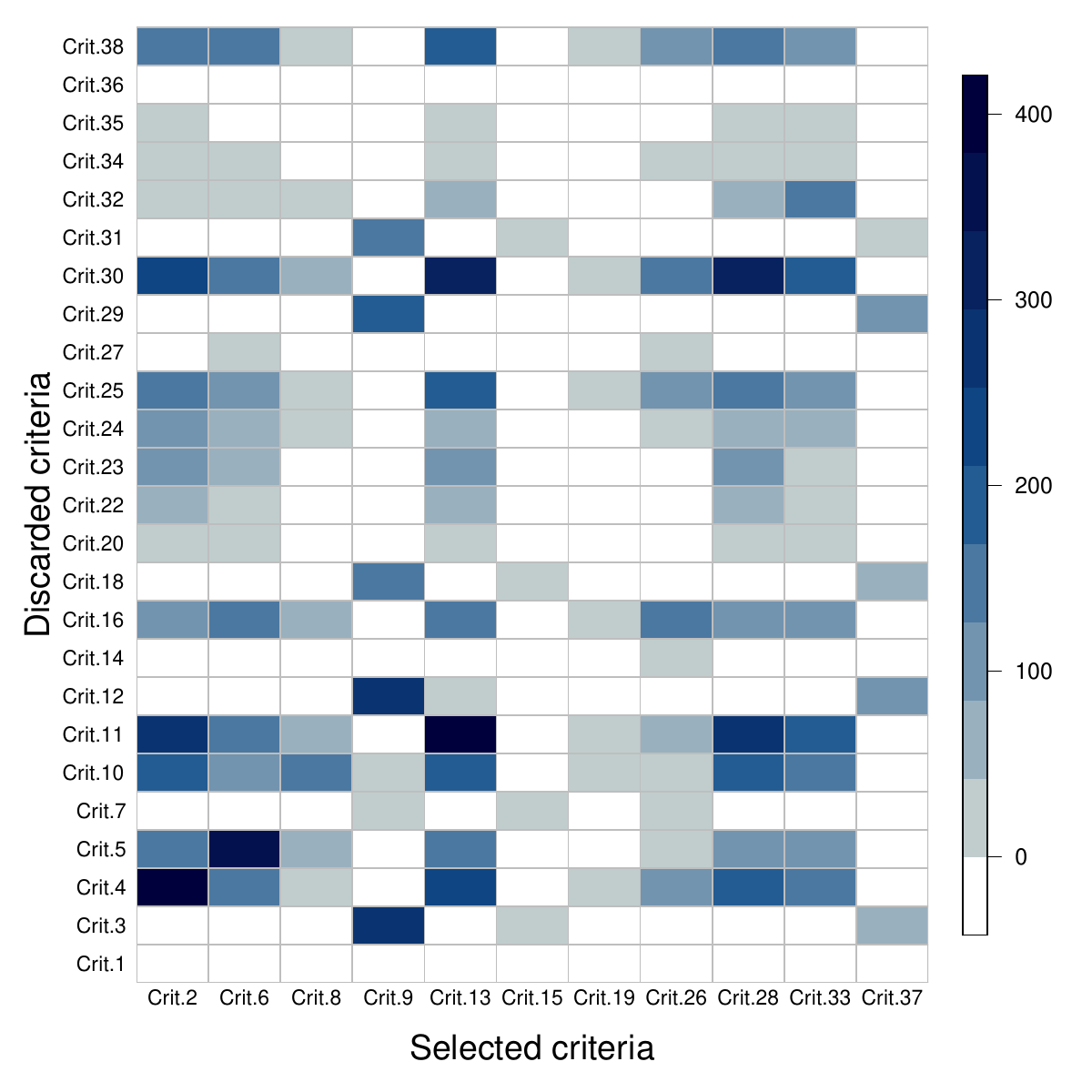}}
\caption{\label{fig:4} Heatmap plot of the BIC difference between the model for the association of each discarded clinical criteria and each selected one and the model for the independence. A white coloured cell indicates no evidence of association between the two variables.}
\end{figure}

We check the association between each discarded clinical criterion and each selected one by calculating the BIC difference in \eqref{eq:5}. The computed differences range from -6.05 to 366.55 and the results are reported in Figure~\ref{fig:4}. Apart from Criterion 1 and Criterion 36, all the discarded criteria present evidence of association with some of the selected criteria. It is also worth to notice that Criterion 36 is the only discarded criterion in the \citet{dean:raftery:2010} modeling framework with the independence assumption between the proposed variable and the clustering ones.

\section{Simulation study}
\label{sim}
In this section we evaluate the proposed variable selection method through two different simulated data scenarios, also discussing the robustness of our method and comparing the results with the \citet{dean:raftery:2010} modelling framework. In both scenarios we simulate 100 datasets for different sample sizes. The scenarios are sketched in Figure~\ref{fig:5} and~\ref{fig:6}. The details of the simulation methodology are exposed in the Supplementary Material, Section 3 [\cite{supplement}].

\begin{figure}
\centering
\begin{minipage}[b]{0.48\linewidth}
\centering
\makebox{\includegraphics[scale=0.9]{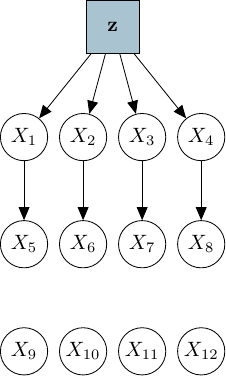}}
\caption{First simulated data scenario.}
\label{fig:5}
\end{minipage}
\begin{minipage}[b]{0.48\linewidth}
\centering
\makebox{\includegraphics[scale=0.9]{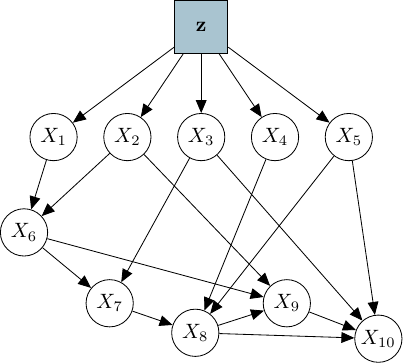}}
\caption{Second simulated data scenario.}
\label{fig:6}
\end{minipage}
\end{figure}    

\subsection{First scenario}
\begin{figure}
\centering
\makebox{\includegraphics[scale=0.7]{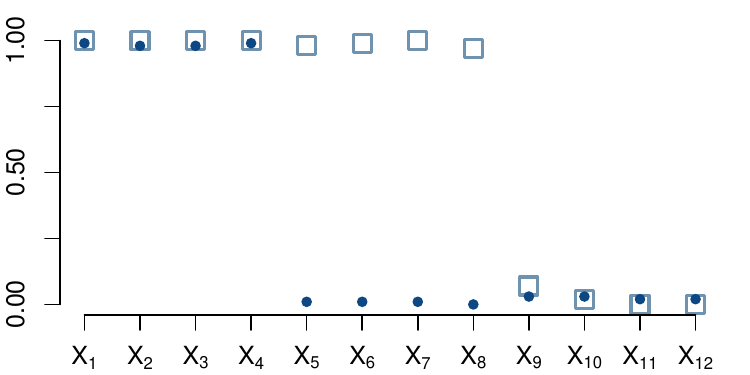}}\\[0.5cm]
\makebox{\includegraphics[scale=0.7]{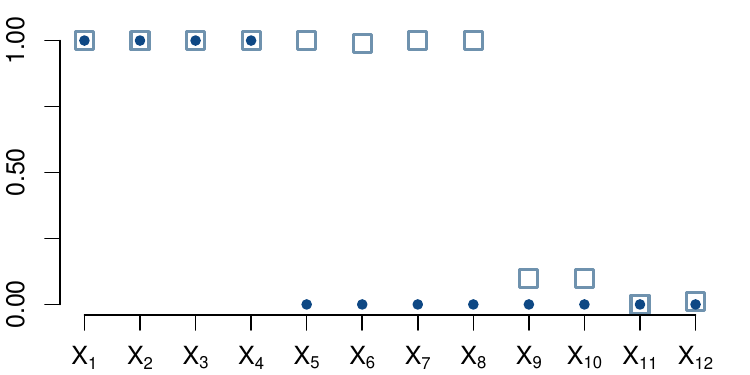}}\\[0.5cm]
\makebox{\includegraphics[scale=0.7]{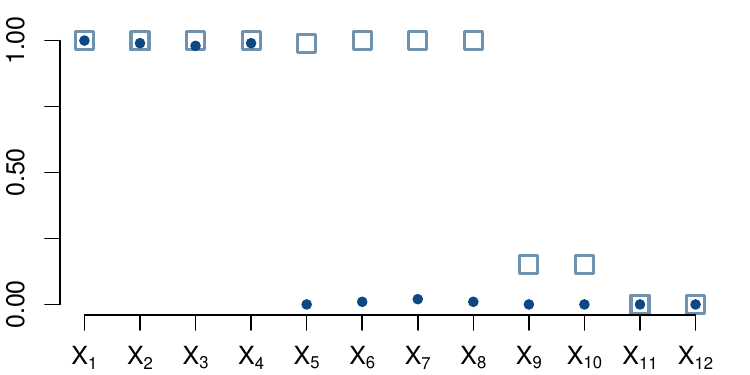}}
\caption{\label{fig:7} First simulation scenario: Proportions of times each variable has been declared a clustering variable by the proposed variable selection method (circle) and the variable selection method with the independence assumption of \citet{dean:raftery:2010} (square). From top: sample sizes corresponding to  500, 750, 1000.}
\end{figure}

\begin{figure}
\centering
\makebox{\includegraphics[scale=0.7]{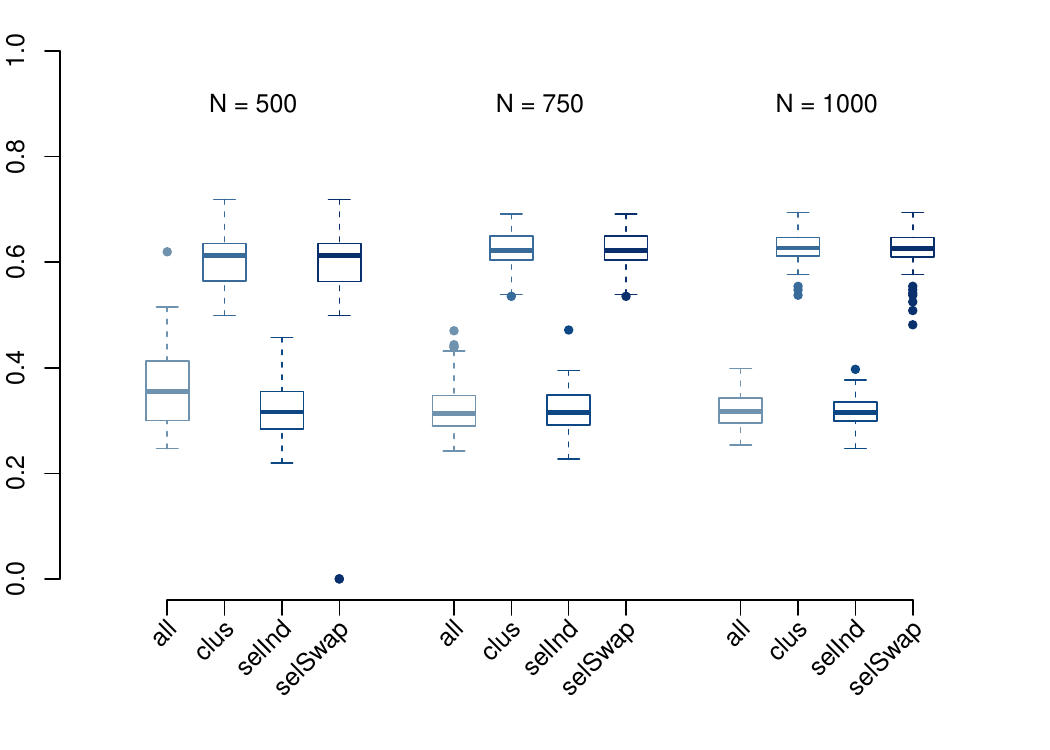}}
\caption{\label{fig:8} First simulation scenario: Boxplots of the ARI between the actual classification of the data and the estimated classification from the LCA  model fitted on: (i) all the variables (\textsf{all}), (ii) the ``true'' clustering variables (\textsf{clus}), (iii) the variables selected by the method with the \citet{dean:raftery:2010} assumption (\textsf{selInd}), (iv) the proposed method (\textsf{selSwap}).}
\end{figure}

\begin{figure}
\centering
\makebox{\includegraphics[scale=0.7]{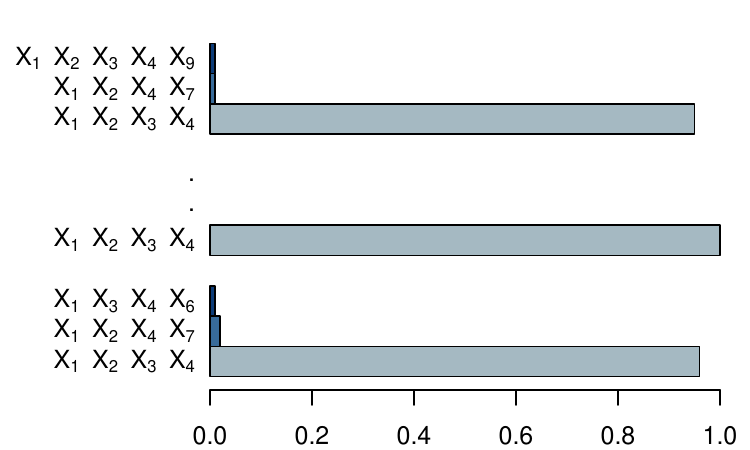}}
\caption{\label{fig:9} First simulation scenario: Proportions of the three most frequent sets of variables declared as relevant for clustering by the presented variable selection method. From top: sample sizes corresponding to  500, 750, 1000.}
\end{figure}

In the first simulation setting we consider 12 categorical random variables. Figure~\ref{fig:5} presents the scenario. Variables $X_1,\,X_2,\,X_3,$ $\,X_4$ are the clustering variables, distributed according to a mixture of $G=3$ multinomial distributions with mixing proportions 0.3, 0.5 and 0.2. Variables $X_5,\,X_6,\,X_7,\,X_8$ are redundant variables, each one generated dependent on one of the clustering variables. The last four variables, $X_9,\,X_{10},\,X_{11},\,X_{12}$ are irrelevant variables not related to the previous ones. We consider three sample sizes: $N= 500, 750, 1000$. Figure~\ref{fig:7} shows the proportion of times each variable was declared a clustering variable by our variable selection method and the variable selection with the independence assumption of \citet{dean:raftery:2010}. Both methods are able to discard the noisy variables. Only the proposed method never selects almost any of the redundant variables, while the \citet{dean:raftery:2010} method includes also the redundant variables in the clustering set, especially as the sample size increases. Figure~\ref{fig:8} displays the boxplots of the ARI between the actual classification of the data and the estimated classification from the LCA model fitted on: (i) all the variables (\textsf{all}), (ii) the ``true'' clustering variables (\textsf{clus}), (iii) the variables selected by the method with the \citet{dean:raftery:2010} assumption (\textsf{selInd}), (iv) the proposed method (\textsf{selSwap}). As expected the inclusion of the redundant variables in the clustering set leads to a poor performance in terms of classification. Figure~\ref{fig:9} presents the three most frequent sets of variables declared as clustering variables by our variable selection procedure. The most selected set is the one composed by the ``true'' clustering variables, and is the only one chosen for a sample size of 750. It is also worth noting that the other selected subsets contain mainly clustering variables.

\subsection{Second scenario}
\begin{figure}
\centering
\makebox{\includegraphics[scale=0.7]{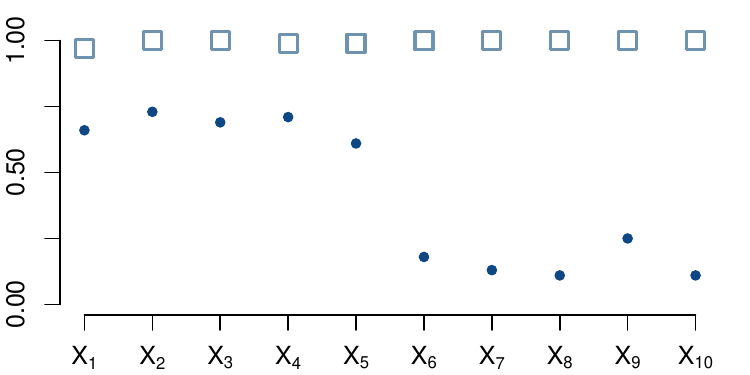}}\\[0.5cm]
\makebox{\includegraphics[scale=0.7]{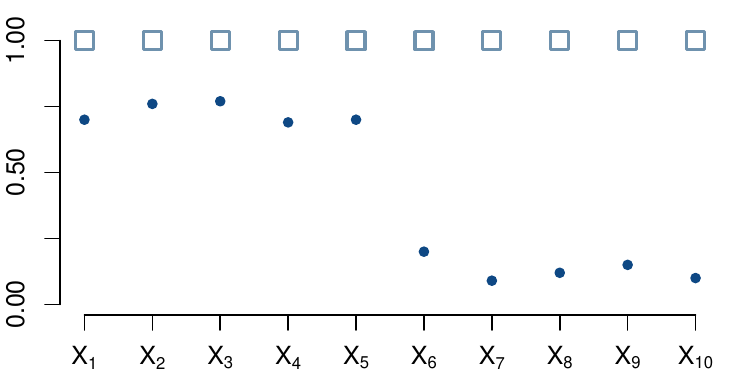}}\\[0.5cm]
\makebox{\includegraphics[scale=0.7]{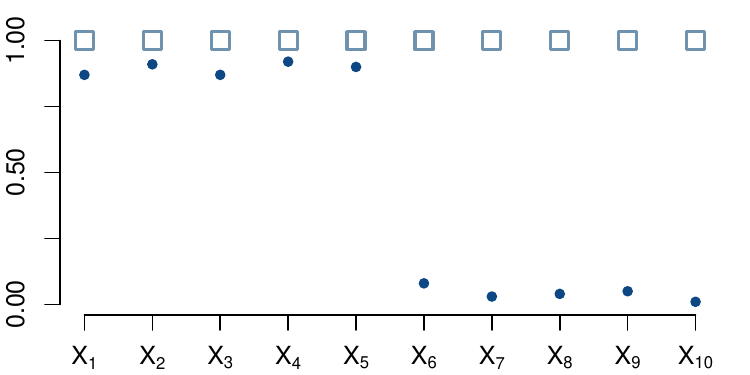}}
\caption{\label{fig:10} Second simulation scenario: Proportions of times each variable has been declared a clustering variable by the proposed variable selection method. From top: sample sizes corresponding to 750, 1000, 1500.}
\end{figure}

\begin{figure}
\centering
\makebox{\includegraphics[scale=0.7]{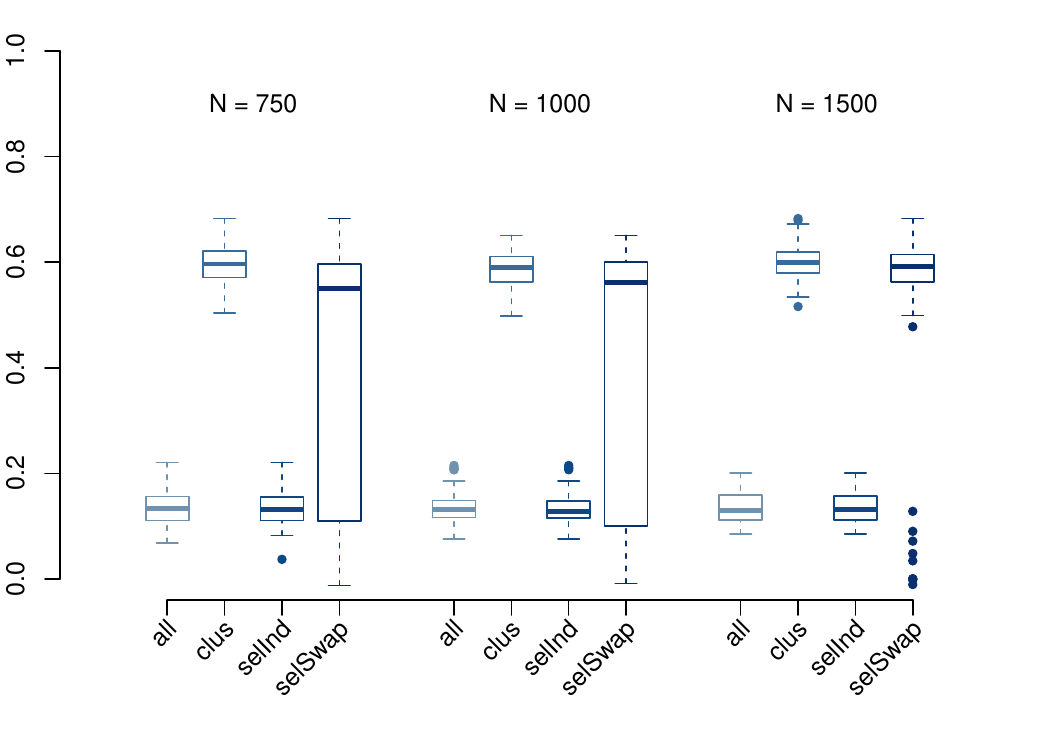}}
\caption{\label{fig:11} Second simulation scenario: Boxplots of the ARI between the actual classification of the data and the estimated classification from the LCA  model fitted on: (i) all the variables (\textsf{all}), (ii) the ``true'' clustering variables (\textsf{clus}), (iii) the variables selected by the method with the \citet{dean:raftery:2010} assumption (\textsf{selInd}), (iv) the proposed method (\textsf{selSwap}).}
\end{figure}

\begin{figure}
\centering
\makebox{\includegraphics[scale=0.7]{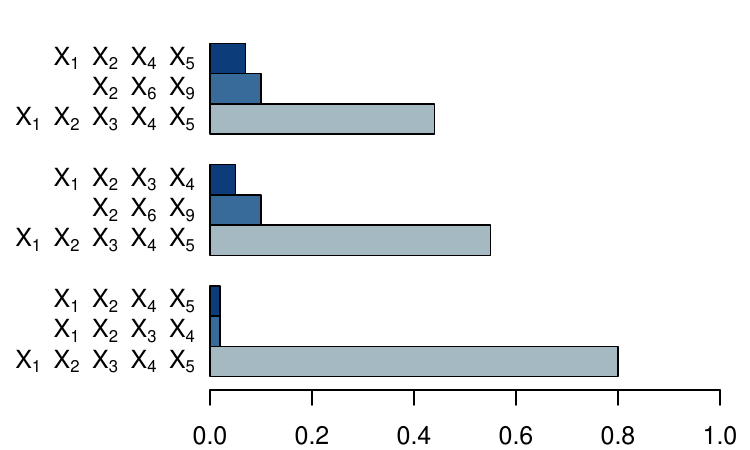}}
\caption{\label{fig:12} Second simulation scenario: Proportions of the three most frequent sets of variables declared as relevant for clustering. From top: sample sizes corresponding to 750, 1000, 1500.}
\end{figure}

In the second simulation setting we consider 10 binary random variables. Figure~\ref{fig:6} shows the scenario. Variables $X_1,\,X_2,\,X_3,\,$ $X_4,\,X_5$ are the clustering variables, distributed according to a mixture of $G=2$ binomial distributions with mixing proportions equal to 0.3 and 0.7. Variables $X_6,\,X_7,\,X_8,\,X_9,\,X_{10}$ are redundant variables; each one of these is generated in order to be dependent on more than one of the clustering variables and the other redundant variables. We consider three sample sizes: $N=750,1000,1500$. Figure~\ref{fig:10} displays the proportion of times each variable was declared a clustering variable. The figure shows that the selection with the \citet{dean:raftery:2010} assumption almost never discard any of the redundant variables. Furthermore with the proposed method the probability of selecting a ``true'' clustering variable increases as $N$ becomes larger. Figure~\ref{fig:11} presents the boxplots of the ARI between the actual classification of the data and the estimated classifications. The classification of the observations based on the selected variables gives on average a better performance in terms of ARI. However there are some situations in which the proposed method does not converge to the selection of the correct set of relevant variables. In Figure~\ref{fig:12} the three most frequent sets declared as clustering variables are shown. Again the set of ``true'' clustering variables is the one selected more often.

\section{Discussion and further work}
\label{disc}
In this paper we have presented an improved variable selection method for LCA that overcomes the limitations of the \citet{dean:raftery:2010} and \citet{white:wyse:2014} methods, which lies in the independence assumption between the selected clustering variables and the variable proposed for removal or inclusion. The proposed method performs the selection of the most informative clustering variables, discarding those that are not informative and those that are redundant. The ability of the method of discriminating among relevant variables and redundant or non-informative variables has been shown in two simulated data settings. 

The work was motivated by the nature of the LBP data examined. In the data all the variables possess good discriminative power, since the clinical criteria list was built by experts in order to best identify the traits of the three classes of pain. The aim was to remove those criteria that are not needed because they contain similar group information to that already included in the selected clinical criteria. This resulted in a smaller set of criteria to be considered in order to derive a mechanisms-based classification of pain. Further, the modeling of the data in an unsupervised manner allowed for the validation of the mechanisms-based classification of pain because the patients clustered into groups that closely correspond to this classification. 

We built the variable selection method on the model comparison framework pioneered by \citet{law:figueiredo:2004} and completely defined in \citet{raftery:dean:2006}. However, another framework for performing variable selection is the regularization approach, although to the authors knowledge it has not been explored yet in categorical data clustering. Furthermore, for continuous data, \citet{celeux:martin:2014} showed that the model comparison approach is a better methodology in terms of classification and variable selection accuracy than the recent regularization method of \citet{witten:tibshirani:2010}.   

We considered a greedy swap-stepwise searching algorithm to perform the variable selection. The idea of replacing a selected variable with one of the discarded variables has already been considered. For example, \citet{miller:2002} in subset selection for regression presents a sequential replacement heuristic where in sequence each of the selected predictors is replaced by one of the non-selected variables. In a model-based clustering context, \citet{tadesse:sha:2005} and \citet{kim:tadesse:2006} use a stochastic search for Bayesian variable selection where the values of a latent variable selection indicator are randomly swapped. Many other searching strategies and metaheuristics could be used in order to conduct a robust search through the solution space and avoid local optima. For example, genetic algorithms \citep{goldberg:1989} have already been applied for variable selection in cluster analysis for market segmentation [\cite{liu:ong:2008}] and subset selection for model-based clustering of continuous data [\cite{scrucca:2016}]. In a high dimensional problem with many variables, a forward algorithm and a headlong search [\cite{badsberg:1992}] can be considered, as has been done in \citet{dean:raftery:2010}. Although in this case the problem of a good initialization of the clustering variables arises. 

The variable selection method is developed in application to clinical criteria selection. However it can be applied to any kind of multivariate categorical data, although its use is limited to only unordered categorical variables. A further extension in that direction is the incorporation of the capability of dealing with ordinal data, which often arise from likert scale questionnaires. In this context it is worth mentioning the work of \citet{arima:2015}, where a Bayesian approach is developed to reduce the items of a questionnaire used to evaluate patients' quality of life, with the goal that the reduced questionnaire will provide the same information of the complete questionnaire. Another limitation of our methodology lies in the local independence assumption of the LCA model. Much work has been done towards relaxing this assumption and allowing class-conditional dependencies between the variables. Among the most recent, \citet{gollini:murphy:2014} present a setting where it is assumed that the class distribution of the categorical variables depends on a number of continuous latent variables, which allow to model the dependences among the observed categorical variables. Another approach is the one by \citet{marbac:biernacki:2015}, where, conditional on a class, the variables are grouped into independent blocks, each one following a specific distribution that takes into account the dependency between variables. Including these frameworks in a variable selection method for clustering categorical data could be promising and may be of interest for further developments.

The variable selection method presented in the paper is implemented in the R [\cite{R}] package \texttt{LCAvarsel}, available on CRAN.

\appendix
\section{Swap-stepwise selection algorithm}\label{app:algo}
Here we give a more detailed description of the swap-stepwise variable selection algorithm for the LCA model. At each stage of the algorithm a greedy search over the model space is conducted and all the variables are examined for being removed, added or swapped. 

Note that in fitting the LCA model we perform multiple runs with random starting values. Also in this case the aim is to allow the search for the global maximum of the log-likelihood rather than a local one; then the model with the greatest log-likelihood is retained. In the following, in the notation we drop the conditioning on the model $\mathcal{M}$ for ease of reading.  

\subsubsection*{Initialization}
Set $G_{\text{max}}$, the maximum number of clusters to be considered for the data. Then when fitting the LCA models, a maximum number $G^*\leq G_{\text{max}}$ of latent classes will be considered at each stage. Here $G^*$ is the maximum number of latent classes that satisfies the identifiability condition in $\eqref{eq:3}$ for the set of variables currently taken into account in fitting the LCA model.

Initialize the set of clustering variables and the set of non-clustering variables by assigning $\mathbf{X}^C=\mathbf{X}$ and $\mathbf{X}^O=\varnothing$ respectively.
\subsubsection*{Removal step}
Fit a LCA model on all the elements contained in the current set of clustering variables $\mathbf{X}^C$, for $1\leq G\leq G^*$ and set 
\[
\text{BIC}_{\text{clus}}=\underset{G}{\text{max}}\bigl\lbrace\,\text{BIC}(\mathbf{X}^C\lvert\mathbf{z})\,\bigr\rbrace.
\]
Then for each variable $X^C_j\in\mathbf{X}^C$ compute
\[
\text{BIC}_{\text{no clus}}(X^C_j) = \underset{G}{\text{max}}\bigl\lbrace\,\text{BIC}(\mathbf{X}_{-j}^C\lvert\mathbf{z})\,\bigr\rbrace + \text{BIC}(X^C_j\lvert\mathbf{X}_j^R \subseteq \mathbf{X}^C_{-j}),
\] 
where $\mathbf{X}^C_{-j} = \mathbf{X}^C\setminus X^C_j$; $\text{BIC}(\mathbf{X}_{-j}^C\lvert\mathbf{z})$ is the BIC of the latent class model on the current clustering variables after removing variable $X^C_j$, maximized over $1\leq G\leq G^*$; $\text{BIC}(X^C_j\lvert\mathbf{X}_j^R \subseteq \mathbf{X}^C_{-j})$ is the BIC of the multinomial logistic regression model of variable $X^C_j$ given the set $\mathbf{X}_j^R$ of selected predictors obtained using the algorithm outlined in the Supplementary Material, Section 2 [\cite{supplement}].

Subsequently for each variable in $\mathbf{X}^C$ estimate the evidence of being a relevant clustering one versus the evidence of not being useful for clustering by computing the difference:
\[
\text{BIC}_{\text{diff}}(X^C_j) = \text{BIC}_{\text{clus}} - \text{BIC}_{\text{no clus}}(X^C_j).
\]
According to the values of $ \text{BIC}_{\text{diff}}(X^C_j)$ rank the current clustering variables in increasing order, generating the ordered set  $\lbrace X^C_{(1)},\, X^C_{(2)},\,\dots,\, X^C_{(M_C)} \rbrace$, with $M_C$ the number of variables in the current set $\mathbf{X}^C$. Then $X^C_{(1)}$ is such that
\[
X^C_{(1)} \,=\,\text{arg}\underset{X^C_j \in \mathbf{X}^C}{\text{min}} \text{BIC}_{\text{diff}}(X^C_j).
\]
Set $X^P = X^C_{(1)}$ and propose it for removal. Next if $\text{BIC}_{\text{diff}}(X^P) < 0$, remove the proposed variable from $\mathbf{X}^C$ and set $\mathbf{X}^C = \mathbf{X}^C\setminus X^P$ and $\mathbf{X}^O = \mathbf{X}^O \cup X^P$; otherwise leave the set of clustering variables unchanged. Go to the swapping step 1.
\subsubsection*{Swapping step 1}
If a variable has been removed in the removal step, set $X_{\text{swap}} = X^C_{(2)}$, otherwise set $X_{\text{swap}} = X^C_{(1)}$.

Swap each variable $X^O_k\in\mathbf{X}^O$ with $X_{\text{swap}}$ generating the sets $\mathbf{X}_k^C = \mathbf{X}^C\setminus X_{\text{swap}} \cup X^O_k$. Fit a LCA model on the set of variables $\mathbf{X}_k^C$ for $1\leq G\leq G^*$ and compute
\[
\text{BIC}_{\text{clus}}(X^O_k)=\underset{G}{\text{max}}\bigl\lbrace\,\text{BIC}(\mathbf{X}_k^C\lvert\mathbf{z})\,\bigr\rbrace + \text{BIC}(X_{\text{swap}}\lvert\mathbf{X}_{\text{swap}}^R \subseteq \mathbf{X}_k^C),
\]
where $\text{BIC}(\mathbf{X}_k^C\lvert\mathbf{z})$ is the BIC of the latent class model on the current clustering variables after swapping the variable $X_{\text{swap}}$ with the variable $X^O_k$, and $\text{BIC}(X_{\text{swap}}\lvert\mathbf{X}_{\text{swap}}^R \subseteq \mathbf{X}_k^C)$ is the BIC of the multinomial logistic regression of variable $X_{\text{swap}}$ given the selected predictors $\mathbf{X}_{\text{swap}}^R$.

Then calculate
\[
\text{BIC}_{\text{no clus}}(X^O_k) = \underset{G}{\text{max}}\bigl\lbrace\,\text{BIC}(\mathbf{X}^C\lvert\mathbf{z})\,\bigr\rbrace + \text{BIC}(X^O_k\lvert\mathbf{X}_k^R \subseteq \mathbf{X}^C),
\] 
where $\text{BIC}(X^O_k\lvert\mathbf{X}_k^R \subseteq \mathbf{X}^C)$ is the BIC of the multinomial logistic regression model of variable $X^O_k$ given the set $\mathbf{X}_k^R$ of relevant predictors in $\mathbf{X}^C$.

Subsequently for each variable in $\mathbf{X}^O$ estimate the evidence of carrying more clustering information than $X_{\text{swap}}$ versus the evidence of containing less clustering information by computing the difference
\[
\text{BIC}_{\text{diff}}(X^O_k) = \text{BIC}_{\text{clus}}(X^O_k) - \text{BIC}_{\text{no clus}}(X^O_k),
\]
and propose for swapping with $X_{\text{swap}}$ the variable $X^P$ such that
\[
X^P \,=\,\text{arg}\underset{X^O_k \in \mathbf{X}^O}{\text{max}} \text{BIC}_{\text{diff}}(X^O_k).
\]
Then if $\text{BIC}_{\text{diff}}(X^P) > 0$, replace $X_{\text{swap}}$ by the proposed variable and set $\mathbf{X}^C = \mathbf{X}^C\setminus X_{\text{swap}} \cup X^P$ and $\mathbf{X}^O = \mathbf{X}^O \setminus X^P  \cup X_{\text{swap}}$; otherwise leave the set of clustering variables unchanged. Go to the inclusion step.
\subsubsection*{Inclusion step}
For each variable $X^O_k\in\mathbf{X}^O$ compute
\[
\text{BIC}_{\text{clus}}(X^O_k) = \underset{G}{\text{max}}\bigl\lbrace\,\text{BIC}(\mathbf{X}_{+k}^C\lvert\mathbf{z})\,\bigr\rbrace,
\]
where $\mathbf{X}_{+k}^C = \mathbf{X}^C \cup X^O_k$; $\text{BIC}(\mathbf{X}_{+k}^C\lvert\mathbf{z})$ is the BIC of the latent class model on the current clustering variables after adding variable $X_k^O$.

Then compute
\[
\text{BIC}_{\text{no clus}}(X^O_k) = \underset{G}{\text{max}}\bigl\lbrace\,\text{BIC}(\mathbf{X}^C\lvert\mathbf{z})\,\bigr\rbrace + \text{BIC}(X^O_k\lvert\mathbf{X}_k^R \subseteq \mathbf{X}^C).
\] 
Subsequently for each variable $X^O_k$ estimate the evidence of being a clustering variable versus the evidence of not being useful for clustering by computing the difference.
\[
\text{BIC}_{\text{diff}}(X^O_k) = \text{BIC}_{\text{clus}}(X^O_k) - \text{BIC}_{\text{no clus}}(X^O_k).
\]
According to the values of $ \text{BIC}_{\text{diff}}(X^O_k)$ rank the current non-clustering variables in decreasing order, generating the ordered set  $\lbrace X^O_{(1)},\, X^O_{(2)},\,\dots,\, X^C_{(M_O)} \rbrace$, with $M_O$ the number of variables in the current set $\mathbf{X}^O$. Then $X^O_{(1)}$ is such that
\[
X^O_{(1)} \,=\,\text{arg}\underset{X^O_j \in \mathbf{X}^O}{\text{max}} \text{BIC}_{\text{diff}}(X^O_j).
\]
Set $X^P = X^O_{(1)}$ and propose it for inclusion in the clustering set. Next if $\text{BIC}_{\text{diff}}(X^P) > 0$, add the proposed variable to $\mathbf{X}^C$ and set $\mathbf{X}^C = \mathbf{X}^C \cup X^P$ and $\mathbf{X}^O = \mathbf{X}^O \setminus X^P$; otherwise leave the set of clustering variables unchanged. Go to the swapping step 2.
\subsubsection*{Swapping step 2}
If a variable has been added in the inclusion step, set $X_{\text{swap}} = X^O_{(2)}$, otherwise set $X_{\text{swap}} = X^O_{(1)}$.

Swap each variable $X^C_j\in\mathbf{X}^C$ with $X_{\text{swap}}$ generating the sets $\mathbf{X}_j^C = \mathbf{X}^C \cup X_{\text{swap}} \setminus X^C_j$. Compute
\[
\text{BIC}_{\text{clus}}(X^C_j)=\underset{G}{\text{max}}\bigl\lbrace\,\text{BIC}(\mathbf{X}^C\lvert\mathbf{z})\,\bigr\rbrace + \text{BIC}(X_{\text{swap}}\lvert\mathbf{X}_{\text{swap}}^R \subseteq \mathbf{X}^C).
\]
Then fit a LCA model on the set of variables $\mathbf{X}_j^C$ for $1\leq G\leq G^*$ and calculate
\[
\text{BIC}_{\text{no clus}}(X^C_j) = \underset{G}{\text{max}}\bigl\lbrace\,\text{BIC}(\mathbf{X}^C_j\lvert\mathbf{z})\,\bigr\rbrace + \text{BIC}(X^C_j\lvert\mathbf{X}_j^R \subseteq \mathbf{X}_j^C),
\]
where $\text{BIC}(\mathbf{X}_j^C\lvert\mathbf{z})$ is the BIC of the latent class model on the current clustering variables after swapping the variable $X_{\text{swap}}$ with the variable $X^C_j$, and $\text{BIC}(X^C_j\lvert\mathbf{X}_j^R \subseteq \mathbf{X}_j^C)$ is the BIC of the multinomial logistic regression model of variable $X^C_j$ given the set $\mathbf{X}_j^R$ of relevant predictors in $\mathbf{X}_j^C$.

Subsequently for each variable in $\mathbf{X}^C$ estimate the evidence of carrying more clustering information than $X_{\text{swap}}$ versus the evidence of containing less clustering information by computing the difference
\[
\text{BIC}_{\text{diff}}(X^C_j) = \text{BIC}_{\text{clus}}(X^C_j) - \text{BIC}_{\text{no clus}}(X^C_j),
\]
and propose for swapping with $X_{\text{swap}}$ the variable $X^P$ such that
\[
X^P \,=\,\text{arg}\underset{X^C_j \in \mathbf{X}^C}{\text{min}} \text{BIC}_{\text{diff}}(X^C_j).
\]
Then if $\text{BIC}_{\text{diff}}(X^P) < 0$, replace $X_{\text{swap}}$ by the proposed variable and set $\mathbf{X}^C = \mathbf{X}^C\setminus X_{\text{swap}} \cup X^P$ and $\mathbf{X}^O = \mathbf{X}^O \setminus X^P  \cup X_{\text{swap}}$; otherwise leave the set of clustering variables unchanged.\\ 

The algorithm starts with two successive removal steps, then it iterates alternating between removal, swapping, inclusion, swapping steps. It stops when all the moves are rejected since no further change can be produced on the set of clustering variables. In the swapping steps we do not look at all possible pairs of variables because it could be too computational demanding. Instead we consider the variable with the largest evidence of being removed or added, because it is the one most likely to be swapped.\\

{\bf Acknowledgments.} We would like to thank the Editor, Associate Editor and Referees whose suggestions and comments helped to improve this paper.

%
\bibliographystyle{apalike}
\bibliography{bibliography.bib}

\newpage

\begin{center}
\textbf{\Huge Supplementary Material}
\vspace*{1cm}
\end{center}

\section{Low back pain data: ``Don't know'' entries}
In the article, patients with ``Don't know'' entries for some clinical indicators were removed for consistency with the approach of \cite{smart:blake:2011}. 

Another reasonable approach would be to consider these entries as \emph{missing at random} \citep{little:2002}, assuming that conditionally on observed and unobserved data, the missingness mechanism does not depend on the unobserved data. Then, for each patient, the framework can be reformulated partitioning the vector of clinical criteria into observed and unobserved components with the result that the missing entries are ignorable for inference about the model; see the approach of \citet{bartolucci:etal:2016} and references therein. Note  that in practice this approach would have the same results as deleting the records corresponding to the missing values.

A further approach would be to add an extra category corresponding to the ``Don't know'' outcome. For some criteria the ``Don't know'' outcome appears only once in the data and adding a extra category would be problematic as only one observation of it would be availabe. However, Criterion 20 contains 27 records labeled as ``Don't know''. In the following we present the results of the variable selection method on the data where an extra category was considered for Criterion 20.

Table~\ref{tab1} contains the results of the variable selection and the latent class model fitting. Using the proposed swap-stepwise variable selection method we retain 8 variables and the BIC selects a 3-class model on these. The selected criteria are 6, 7, 12, 13, 19, 26, 30, 33. With the \citet{dean:raftery:2010} approach and the standard stepwise selection different number of classes and larger sets of clustering variables are selected. 

Table~\ref{crosstab1} presents a cross-tabulation of the estimated partition on the 8 selected variables versus the expert-based classification of the patients. Again, it seems quite reasonable to match the three detected classes to the three mechanisms-related types of pain.

Table~\ref{crosstab2} presents a cross-tabulation of the partitions estimated by the variable selection method applied on the two sets of data. The table compares the estimated partition for the matching pairs of observations between the two sets of data.

\newpage

\begin{table}
\centering
\caption{\label{tab1} Clustering summary of the LCA model for different sets of variables and different number of classes for the low back pain data with an extra category considered for Criterion 20}
\begin{tabular}{llcrr}
\toprule
\begin{tabular}{@{}l@{}}\textbf{Selection}\\\textbf{method}\end{tabular} & \textbf{Variables}	&\begin{tabular}{@{}l@{}}\textbf{N. latent}\\\textbf{classes}\end{tabular} & \textbf{BIC} & \textbf{ARI}\\
\midrule
--	&	All		& 5		& -13539.36	& 0.49\\
--	&	All		& $3^*$		& -13756.95	& 0.79\\
Dean and Raftery	&	35 Criteria	& 5		& -13047.50	& 0.48\\
Stepwise		&	12 Criteria	& 4		& -4558.85	& 0.60\\
Swap-stepwise		&	8 Criteria	& 3		& -3151.30	& 0.62\\
\bottomrule
\end{tabular}\\
{\footnotesize{$^*$ The number of classes was fixed to this value in advance.}}
\end{table}

\vspace*{1cm}

\begin{table}[hb]
\caption{\label{crosstab1} Cross-tabulation between the estimated partition on the 8 clustering variables and the expert-based classification of the patients.}
\centering
\begin{tabular}{llccc}
\toprule
& & \multicolumn{3}{c}{\textbf{\emph{Estimated}}}\\
& & \textbf{Class 1} & \textbf{Class 2} & \textbf{Class 3} \\
\multirow{3}{4em}{\emph{\textbf{Expert-based}}} & \textbf{Nociceptive} & 236 &   13 &  3 \\ 
& \textbf{Peripheral Neuropathic} &   38 &   64 &  0 \\ 
& \textbf{Central Sensitization} &   7 &  2 &   89 \\ 
\bottomrule
\end{tabular}
\end{table}

\vspace*{1cm}

\begin{table}[hb]
\caption{\label{crosstab2} Cross-tabulation between two estimated partitions: the columns contains the partition estimated on the data with an extra category for Criterion 20; the rows contain the partition estimated on the data analysed in the article. Note that only matching pairs of observations are considered.}
\centering
\begin{tabular}{llccc}
\toprule
& & \multicolumn{3}{c}{\textbf{\emph{``Don't know'' included}}}\\
& & \textbf{Class 1} & \textbf{Class 2} & \textbf{Class 3} \\ & \textbf{Class 1} & 208 &   9 &  1 \\ 
& \textbf{Class 2} &   48 &   63 &  1 \\ 
& \textbf{Class 3} &   6 &  3 &   86 \\ 
\bottomrule
\end{tabular}
\end{table}

\newpage

\section{Backward stepwise selection in regression}
\label{appendix:a}
In the following we describe the backward stepwise selection algorithm used to choose the set of predictors $\mathbf{X}^R$ of the proposed variable in the multinomial logistic regression model.

\begin{itemize}
\item \textbf{Initialization}\\ Initialize the set of relevant predictors $\mathbf{X}^R$ and the set of not relevant ones $\mathbf{X}^O_{\text{reg}}$ by assigning $\mathbf{X}^R=\mathbf{X}^C$ and $\mathbf{X}^O_{\text{reg}}=\varnothing$ (note that at each stage $\mathbf{X}^R\cup\mathbf{X}^O_{\text{reg}}=\mathbf{X}^C$).

\item \textbf{Removal step}\\ For all the variables $X_j$ in $\mathbf{X}^R$ compute
\[
\text{BIC}_{\text{diff reg}}(X_j) = \text{BIC}(X_m \lvert \mathbf{X}^R) - \text{BIC}(X_m \lvert \mathbf{X}^R_{-j}),
\]
where $\mathbf{X}^R_{-j}=\mathbf{X}^R\setminus X_j$, $\text{BIC}(X_m \lvert \mathbf{X}^R)$ is the BIC of the regression of variable $m$ on the current set of relevant predictors and $\text{BIC}(X_m \lvert \mathbf{X}^R_{-j})$ is the BIC of the regression after removing variable $j$. 

Then propose to be removed from $\mathbf{X}^R$ the variable $X^P_{\text{reg}}$ such that
\[
X^P_{\text{reg}} \,=\,\text{arg}\underset{X_j \in \mathbf{X}^R}{\text{min}} \text{BIC}_{\text{diff reg}}(X_j), 
\]
and if $\text{BIC}_{\text{diff reg}}(X^P_{\text{reg}}) \leq 0$ remove it and set $\mathbf{X}^R=\mathbf{X}^R\setminus X^P_{\text{reg}}$ and $\mathbf{X}^O_{\text{reg}}=\mathbf{X}^O_{\text{reg}}\cup X^P_{\text{reg}}$; otherwise leave the set of current relevant predictors unchanged and go to the inclusion step.

\item \textbf{Inclusion step}\\ For all the variables $X_k$ in $\mathbf{X}^O_{\text{reg}}$ compute
\[
\text{BIC}_{\text{diff reg}}(X_k) = \text{BIC}(X_m \lvert \mathbf{X}_{+k}^R) - \text{BIC}(X_m \lvert \mathbf{X}^R),
\]
where $\mathbf{X}^R_{+k}=\mathbf{X}^R\cup X_k$ and $\text{BIC}(X_m \lvert \mathbf{X}_{+k}^R)$ is the BIC of the regression of variable $m$ on the current set of relevant predictors after adding variable $k$. 

Then propose to be included into $\mathbf{X}^R$ the variable $X^P_{\text{reg}}$ such that
\[
X^P_{\text{reg}} \,=\,\text{arg}\underset{X_k \in \mathbf{X}^O_{\text{reg}}}{\text{max}} \text{BIC}_{\text{diff reg}}(X_k), 
\]
and if $\text{BIC}_{\text{diff reg}}(X^P_{\text{reg}}) > 0$ include it and set $\mathbf{X}^R=\mathbf{X}^R\cup X^P_{\text{reg}}$ and $\mathbf{X}^O_{\text{reg}}=\mathbf{X}^O_{\text{reg}}\setminus X^P_{\text{reg}}$; otherwise leave the set of current relevant predictors unchanged and go to the removal step.
\end{itemize}

The algorithm starts with two successive removal steps, then it iterates alternating between removal and inclusion step. It stops when the set of relevant predictors remains unchanged after consecutive removal and inclusion steps.
\newpage

\section{Details of the simulated data experiments}
\label{appendix:b}

\subsection{First simulation setting}
In the first simulated data experiment we considered 12 variables. Variables $X_1,\,\dots,\,X_4$ are the relevant clustering variables, $X_5,\,\dots,\,X_8$ are the redundant variables and $X_9,\,\dots,\,X_{12}$ are noisy variables. We simulated data from a 3-class model on the clustering variables. The actual parameters for the relevant variables and the noisy variables are given in Table~\ref{appendixB:tab:1}.

We considered each of the redundant variables to be correlated only to one of the clustering variables. To generate them we used the following transitions matrices:
\[
P_{1,5} = \left[
\begin{array}{cc}
0.90 & 0.10\\
0.20 & 0.80
\end{array} \right] \quad
P_{2,6} = \left[
\begin{array}{ccc}
0.70 & 0.10 & 0.20\\
0.20 & 0.70 & 0.10\\
0.10 & 0.10 & 0.80
\end{array} \right] \quad
\]
\[
P_{3,7} = \left[
\begin{array}{ccc}
0.80 & 0.10 & 0.10\\
0.05 & 0.90 & 0.05\\
0.20 & 0.10 & 0.70
\end{array} \right] \quad
P_{4,8} = \left[
\begin{array}{cccc}
0.70 & 0.10 & 0.10 & 0.10\\
0.10 & 0.80 & 0.05 & 0.05\\
0.10 & 0.20 & 0.60 & 0.10\\
0.05 & 0.10 & 0.05 & 0.80
\end{array} \right]
\]
Thus each redundant variable is simulated from the corresponding clustering variable by means of the relative transition matrix, i.e. $X_5$ is generated from $X_1$ using $P_{1,5}$ and so on. Each cell $p_{ij}$ of the matrices contains the conditional probability that the redundant variable assumes a category $j$ given that the corresponding relevant variable assumed value $i$; for example, for matrix $X_{2,6}$, $\Pr(X_6 = 1\lvert x_2 = 1) = 0.70$, $\Pr(X_6 = 2\lvert x_2 = 1) = 0.10$, and so on. Then the categories of the correlated variables are sampled using these conditional probabilities. Note that the values in the diagonal express the degree of correlation between the clustering variable and the redundant one (the larger the value, the larger is the correlation).

\subsection{Second simulation setting}
In the second setting we generated 10 binary variables (with categories 1 and 2). Variables $X_1,\,\dots,\,X_5$ are the relevant clustering variables, $X_6,\,\dots,\,X_{10}$ are the redundant variables. We considered a 2-class model on the clustering variables, with mixing proportions equal to 0.7 and 0.3. Table~\ref{appendixB:tab:2} lists the actual class conditional probabilities of occurrence for the clustering variables.

This time we consider the following transition matrix
\[
P = \left[
\begin{array}{cc}
0.80 & 0.20\\
0.20 & 0.80
\end{array} \right]
\]
Let now $Y_{j}$ the intermediate variable simulated from variable $j$ through $P$ and used to generate a redundant variable. Then the redundant variables are defined as follows
\begin{eqnarray}
X_6 &=& 2 \quad \text{if} \quad Y_1 + Y_2 \geq 3, \nonumber\\
X_7 &=& 2 \quad \text{if} \quad Y_3 + Y_6 \geq 4, \nonumber\\
X_8 &=& 2 \quad \text{if} \quad Y_4 + Y_5 + Y_7 \geq 5, \nonumber\\
X_9 &=& 2 \quad \text{if} \quad Y_2 + Y_6 + Y_8 \geq 5, \nonumber\\
X_{10} &=& 2 \quad \text{if} \quad Y_3 + Y_5 + Y_8 + Y_9 \leq 5, \nonumber
\end{eqnarray}
and each variable assumes category 1 if otherwise.

\clearpage

\begin{table}[!h]
\caption{\label{appendixB:tab:1} Actual parameters for the clustering variables and the noisy variables in the first simulation setting. For each class the corresponding mixing proportion is reported in brackets.}
\centering
\begin{tabular}{llccc}
\toprule
\textbf{Variable} & \textbf{Category} & \textbf{Class 1} & \textbf{Class 2} & \textbf{Class 3} \\
	     &		    &	   (0.3)  &     (0.5)	&      (0.2)  \\ 
\midrule
$X_1$ & 1 & 0.1 & 0.3 & 0.8 \\ 
      & 2 & 0.9 & 0.7 & 0.2 \\[5pt]
$X_2$   & 1 & 0.1 & 0.2 & 0.8 \\ 
  	& 2 & 0.1 & 0.6 & 0.1 \\ 
  	& 3 & 0.8 & 0.2 & 0.1 \\[5pt]
$X_3$   & 1 & 0.1 & 0.7 & 0.2 \\ 
   	& 2 & 0.7 & 0.1 & 0.2 \\ 
   	& 3 & 0.2 & 0.2 & 0.6 \\[5pt]
$X_4$  & 1 & 0.7 & 0.1 & 0.2 \\ 
       & 2 & 0.1 & 0.1 & 0.1 \\ 
       & 3 & 0.1 & 0.7 & 0.1 \\ 
       & 4 & 0.1 & 0.1 & 0.6 \\[5pt]
$X_9$ & 1 & 0.7 & 0.7 & 0.7 \\ 
      & 2 & 0.3 & 0.3 & 0.3 \\[5pt] 
$X_{10}$ & 1 & 0.6 & 0.6 & 0.6 \\ 
 	 & 2 & 0.4 & 0.4 & 0.4 \\[5pt]
$X_{11}$   & 1 & 0.4& 0.4 & 0.4 \\ 
 	   & 2 & 0.3 & 0.3 & 0.3 \\ 
  	   & 3 & 0.3 & 0.3 & 0.3 \\[5pt] 
$X_{12}$   & 1 & 0.2 & 0.2 & 0.2 \\ 
  	   & 2 & 0.3 & 0.3 & 0.3 \\ 
  	   & 3 & 0.5 & 0.5 & 0.5 \\
\bottomrule
\end{tabular}
\end{table}

\begin{table}[!h]
\caption{\label{appendixB:tab:2} Actual class conditional probabilities of occurrence for the clustering variables of the second simulation experiment.}
\centering
\begin{tabular}{lcc}
\toprule
\textbf{Variable} &  \textbf{Class 1} & \textbf{Class 2} \\
\midrule
$X_1$ & 0.4 & 0.8 \\ 
$X_2$ & 0.8 & 0.4 \\ 
$X_3$ & 0.2 & 0.5 \\ 
$X_4$ & 0.1 & 0.8 \\ 
$X_5$ & 0.6 & 0.3 \\
\bottomrule
\end{tabular}
\end{table}
\vfill
\newpage

\section{Clinical criteria checklist}
\vspace*{-0.5cm}
\begin{table}[!h]
\hspace*{-0.8cm}
 \footnotesize
 \centering
 \begin{tabular}{P{1.8cm}p{14.5cm}}
 \toprule
 \textbf{Criterion}	&	\textbf{Description}\\
 \midrule
1.	& Pain of recent onset\\
2.	& Pain associated with and in proportion to trauma, a pathologic process or movement/postural dysfunction\\
3.	& History of nerve injury, pathology, or mechanical compromise\\
4.	& Pain disproportionate to the nature and extent of injury or pathology\\
5.	& Usually intermittent and sharp with movement/mechanical provocation; may be a more constant dull ache or throb at rest\\
6.	& More constant/unremitting pain\\
7.	& Pain variously described as burning, shooting, sharp, or electric-shock-like\\
8.	& Pain localized to the area of injury/dysfunction (with/without some somatic referral)\\
9.	& Pain referred in a dermatomal or cutaneous distribution\\
10.	& Widespread, nonanatomic distribution of pain\\
11.	& Clear, proportionate mechanical/anatomic nature to aggravating and easing factors\\
12.	& Mechanical pattern to aggravating and easing factors involving activities/postures associated with movement, loading, or compression of neural tissue\\
13.	& Disproportionate, nonmechanical, unpredictable pattern of pain provocation in response to multiple/nonspecific aggravating/easing factors\\
14.	& Reports of spontaneous (ie, stimulus-independent) pain and/or paroxysmal pain (ie, sudden recurrences and intensification of pain)\\
15.	& Pain in association with other dysesthesias (eg, crawling, electrical, heaviness)\\
16.	& Pain of high severity and irritability (ie, easily provoked, taking longer to settle)\\
17$^*$.	& Pain in association with other symptoms of inflammation (ie, swelling, redness, heat)\\
18.	& Pain in association with other neurological symptoms (eg, pins and needles, numbness, weakness)\\
19.	& Night pain/disturbed sleep\\
20.	& Responsive to simple analgesia/NSAIDs\\
21$^*$.	& Less responsive to simple analgesia/NSAIDs and/or more responsive to antiepileptic (eg, Lyrica)/antidepression (eg, Amitriptyline) medication\\
22.	& Usually rapidly resolving or resolving in accordance with expected tissue healing/pathology recovery times\\
23.	& Pain persisting beyond expected tissue healing/pathology recovery times\\
24.	& History of failed interventions (medical/surgical/therapeutic)\\
25.	& Strong association with maladaptive psychosocial factors (eg, negative emotions, poor self-efficacy, maladaptive beliefs, and pain behaviors, altered family/work/social life, medical conflict)\\
26.	& Pain in association with high levels of functional disability\\
27.	& Antalgic (ie, pain relieving) postures/movement patterns\\
28.	& Clear, consistent, and proportionate mechanical/anatomic pattern of pain reproduction on movement/mechanical testing of target tissues\\
29.	& Pain/symptom provocation with mechanical/movement tests (eg, Active/Passive, Neurodynamic, ie, SLR) that move/load/compress neural tissue\\
30.	& Disproportionate, inconsistent, nonmechanical/nonanatomic pattern of pain provocation in response to movement/mechanical testing\\
31.	& Positive neurological findings (altered reflexes, sensation, and muscle power in a dermatomal/myotomal or cutaneous nerve distribution)\\
32.	& Localized pain on palpation\\
33.	& Diffuse/nonanatomic areas of pain/tenderness on palpation\\
34.	& Positive findings of allodynia within the distribution of pain\\
35.	& Positive findings of hyperalgesia (primary and/or secondary) within the distribution of pain\\
36.	& Positive findings of hyperpathia within the distribution of pain\\
37.	& Pain/symptom provocation on palpation of relevant neural tissues\\
38.	& Positive identification of various psychosocial factors (eg, catastrophization, fear-avoidance behavior, distress)\\
\bottomrule
 \end{tabular}
 {\footnotesize{$^*$ Criteria 17 and 21 were not available in the data analysed in the article.}}
\end{table}
\newpage

\section{Notation}
 \begin{tabular}{L{2.8cm}L{14.5cm}}
  $N$			&	Number of observations (number of patients in the article)\\
  $M$			&	Number of variables (number of clinical criteria in the article)\\
  $G$			&	Number of clusters/groups\\
  $C_m$			&	Number of categories of variable $m$\\
  $n = 1,\dots,N$	&	Observations subscript\\
  $m = 1,\dots,M$	&	Variables subscript\\
  $g = 1,\dots,G$	&	Classes subscript\\
  $c = 1,\dots,Cm$	&	Categories subscript for variable $m$\\
  $\mathbf{X}$		&	Data matrix of dimension $N\times M$\\
  $\mathbf{X}_n$	&	Single observation as a $M$-dimensional vector\\
  $X_m$			&	Variable $m$\\
  $\mathbf{z}$		&	Cluster membership indicator variable\\
  $\mathbf{z}_n$	&	Cluster memeberhip indicator variable for observation $n$\\
  $\bm{\theta}_g$	&	Parameter set of component $g$\\
  $\theta_{gmc}$	&	Probability of variable $m$ taking value $c$ within class $g$\\
  $\mathcal{M}_{A}$	&	Model $A$\\
  $\mathbf{X}^C$	&	Set of clustering variables\\
  $X^P$			&	Variable proposed for adding/removal\\
  $\mathbf{X}^O$	&	Set of non-clustering variables\\
  $\mathbf{X}^R$	&	Set of relevant predictors in the regression step\\
 \end{tabular}

\vspace*{1cm}

\end{document}